\documentclass{aa}  

\usepackage{graphicx}
\usepackage{xcolor}
\usepackage{txfonts}
\usepackage{placeins}
\defcitealias{VelizAstudillo+2024}{Paper I}

\begin{document}

   \title{The effect of dynamical states on galaxy cluster populations}

   \subtitle{II. Comparison of galaxy properties and fundamental relations}

   \author{S. V\'eliz Astudillo \inst{1} \and
   E. R. Carrasco \inst{2} \and
   J. L. Nilo Castell\'on \inst{1} \and
   A. Zenteno \inst{3} \and
   H. Cuevas \inst{1}}
   \institute{Departamento de Astronom\'ia, Universidad de La Serena, Av. Ra\'ul Bitr\'an 1305, La Serena, Chile \and
   International Gemini Observatory, NFS NOIRLab, Casilla 603, La Serena, Chile \and
   Cerro Tololo Inter-American Observatory, NFS NOIRLab, Casilla 603, La Serena, Chile}

   \date{Received September 15, 1996; accepted March 16, 1997}
 
  \abstract
   {Galaxy clusters provide a unique environment in which to study the influence of extreme conditions on galaxy evolution. The role of cluster dynamical states in shaping the physical and morphological properties of member galaxies remains an open question.}
   {We aim to assess the impact of the dynamical state of massive ($M_{500} \geq 1.5 \times 10^{14} M_{\odot}$) galaxy clusters on the physical and structural properties of their member galaxies, and also in their fundamental relations in the redshift range $0.10 < z < 0.35$, comparing relaxed and disturbed clusters.}
   {We use a mass-matched sample of galaxies from relaxed and disturbed clusters. Morphological types were assigned using both parametric and nonparametric methods, while physical properties were derived through spectral energy distribution fitting. Galaxies were further divided into subpopulations based on their colors and stellar masses to investigate trends with cluster dynamical states. Additionally, we examined correlations between the dynamical state of clusters and their fundamental relations, such as color-magnitude, mass-size, morphology-density, and SF-density relations.}
   {The dynamical state of galaxy clusters does not alter their fundamental relations at low redshift, nor does it significantly affect the mean or dispersion of galaxy properties. However, it does impact the distributions at the level of third- and fourth-order moments, introducing asymmetries and heavier tails in the structural parameters and specific star formation rates (sSFRs) of galaxies. The greatest effects are observed in the sSFRs of low-mass and red sequence galaxies.}
   {These findings suggest that, at low redshift, the fundamental relations of massive galaxy clusters are already well established and resilient to recent dynamical activity. Nonetheless, the influence of the dynamical state on the higher-order moments of galaxy properties indicates that environmental processes associated with disturbed clusters -- such as tidal interactions, ram-pressure stripping, and harassment -- still leave measurable imprints, particularly on low-mass and red sequence galaxies. This is consistent with the idea that galaxy evolution is shaped both by early pre-processing and by subsequent interactions within dynamically active environments.
}


   \maketitle

\section{Introduction}
\label{sec:introduction}

Understanding the formation and evolution of galaxies remains a key challenge in astrophysics. While it is well established that galaxies undergo secular evolution primarily driven by their intrinsic properties \citep[e.g.,][]{Kormendy+Kennicutt2004}, there is also strong evidence that galaxy properties are closely related to their environment \citep[e.g.,][]{Kauffmann+04,Blanton&Moustakas09,Peng+2010}. These environmental trends have been primarily parametrized through the morphology-density, morphology-clustercentric radius, and star formation-density relations \citep[e.g.][]{Dressler80,Dressler+85,Whitmore&Gilmore91, Dressler+97,Kauffmann+2004}, in which the fraction of spiral or star-forming galaxies decreases with local density. Evidence of these relations has been observed from the local Universe up to $z \sim 2$ \citep[e.g.,][]{Postman+05,Sazonova+20}, and they have also been reproduced in simulations based on the standard $\Lambda$ cold dark matter ($\Lambda$CDM) cosmological model \citep[e.g.,][]{Teklu+17,Pfeffer+23}.

Behind these observations and model predictions, several physical processes have been proposed in the literature to explain the trends between galaxies and their environment. Among the most notable are tidal stripping \citep{Byrd&Valtonen90,Valluri93}, galaxy harassment \citep{Moore+96}, galaxy mergers, starvation \citep{Larson+80}, and ram pressure stripping \citep{Gunn&Gott72}. For a comprehensive review of these and other physical processes and their applications in different environments, see \citet{Boselli&Gavazzi06,Mo+2010}.

Given their nature, as the densest structures in the Universe, galaxy groups and clusters are ideal environments for studying galaxy evolution and the potential physical processes affecting them, thereby producing the observed trends with local density. Indeed, this has been the most extensively developed approach to date \citep[e.g.,][]{Butcher+Oemler1984,Balogh+99,Poggianti+99,Treu+2003,Carrasco+2010,NiloCastellon+14,LimaDias+24}. However, under the hierarchical paradigm of structure formation, galaxy clusters are also subject to interactions with other clusters and the accretion of smaller structures \citep[e.g.,][]{Voit+05,Molnar16}. In fact, this is how they assemble, and when observing different galaxy clusters, they can be found in various dynamical states \citep[e.g.,][]{DresslerShectman88,Santos+08,FakhouriMa10,Skillman+13,Golovich+19,zenteno25}.

Following this idea, and considering that galaxy cluster mergers are the second most energetic events after the Big Bang, releasing energy on the order of 10$^{64}$ erg/s \citep{Sarazin02,KravstovBorgani12}, the properties of galaxies in clusters with different dynamical states have been studied under the hypothesis that these extreme environments can affect their evolution. An example of this is the study by \citet{Ribeiro+13}, which found that the fraction of red galaxies as a function of clustercentric distance does not show significant differences between relaxed and disturbed clusters. However, they found that the luminosity function of relaxed clusters has a steeper faint end slope than that of disturbed clusters at low redshift ($z \leq 0.1$). In contrast, \citet{Zenteno+20} found that while relaxed and disturbed clusters show differences at the faint end slope of the luminosity function, this occurs at a higher redshift ($z \gtrsim 0.5$), with disturbed clusters exhibiting a steeper slope. Thus, \citet{Zenteno+20} hypothesized that at $z \gtrsim 0.5$, galaxy populations begin to show differences between relaxed and disturbed clusters, which was further corroborated by \citet{Aldas+23} in terms of the color-magnitude relation of these systems.

The present work corresponds to the second article of a two-part series aimed at investigating the impact of the dynamical state of massive clusters ($M_{500} \geq 1.5 \times 10^{14}$ M$_{\odot}$) on the physical and morphological properties of their member galaxies, within the redshift range $0.10 < z < 0.35$. In \citet[][hereafter Paper I]{VelizAstudillo+2024}, we detail the selection of the cluster and member galaxy sample, as well as determine the dynamical states of the galaxy clusters using X-ray images from the Chandra \citep{Weisskopf+2000} and XMM-Newton \citep{Jansen+2001} archives, together with optical images from the DESI Legacy Imaging Survey Data Release 10. In this second paper, we present the calculation of the physical and morphological properties of galaxies, as well as the analysis of the impact of the dynamical states on galaxy populations in the most relaxed and disturbed clusters.

This paper is organized as follows. In Section \ref{sec:data}, we present the description of the data. In Section \ref{sec:galaxy-properties}, we detail the methods used to determine the physical and structural properties of galaxies, and the morphological classification is explained in Section \ref{sec:morphological-classification}. In Section \ref{sec:results}, we present the results in a comparative context between relaxed and disturbed clusters. In Section \ref{sec:discussion}, we discuss the results, and finally, we summarize the work and present the conclusions in Section \ref{sec:conclusions}.

Throughout this work, we adopt a flat $\Lambda$CDM cosmology, assuming $H_0 = 69.3$ km s$^{-1}$, $\Omega_{\Lambda} = 0.721$, and $\Omega_{m} = 0.287$ \citep[WMAP-9,][]{Hinshaw+13}. Unless specifically stated, the magnitudes used in this article are quoted in the AB system.

\section{Data}
\label{sec:data}

\subsection{Cluster sample}
\label{sec:cluster-sample}

We used a sample of 87 massive galaxy clusters ($M_{500} \geq 1.5 \times 10^{14} M_{\odot}$) defined in \citetalias{VelizAstudillo+2024}. These clusters are located within the redshift range $0.10 < z < 0.35$ and have already been classified by their dynamical state into relaxed, intermediate, and disturbed clusters. In this work, we focus exclusively on relaxed and disturbed clusters, representing the most extreme cases in the sample. Galaxy members were assigned using photometric redshifts. To ensure homogeneity, we applied a magnitude limit with a cut of $m_r < m_r^* + 3$, where $m_r^*$ is the characteristic magnitude of the cluster in the $r$ band. Furthermore, we restricted our study to within $R_{200}$ for each cluster and verified that galaxies are homogeneously distributed within this range, i.e., the fraction of galaxies is roughly the same at different clustercentric distance bins (see Fig. 3 in \citetalias{VelizAstudillo+2024}).

\subsection{Images}
\label{sec:images}

We used images from the DESI Legacy Imaging Survey Data Release 10 (LS DR10). Although this survey combines data from three different surveys to cover more than 20,000 squares degrees in the sky \citep{Dey+19}, our entire cluster sample is contained within the Dark Energy Camera Legacy Survey \citep[DECaLS,][]{Blum+16} footprint. This survey provides imaging data in the $griz$ bands, reaching a uniform depth of $5\sigma$ at $m_g = 24.7$, $m_r = 23.9$, $m_i = 23.6$, and $m_z = 23.0$ for point sources.

\subsection{Galaxy catalogs}
\label{sec:catalogs}

On the one hand, we used photometric catalogs from the LS DR10. These catalogs were generated using the software \texttt{The Tractor} \citep{Lang+16}. This code employs a probabilistic method to fit surface brightness models to sources detected in astronomical images. The models can be point sources (PSF), round exponential galaxies with variable radius (REX), de Vaucouleurs (DEV) profiles, exponential (EXP) profiles, and S\'ersic (SER) profiles, and these are fitted through a $\chi^2$ minimization problem. It is worth noting that in addition to optical data, the catalog also includes infrared information from the WISE five-year co-added image, also known as unWISE \citep{Wright+10}, in the $W_1$, $W_2$, $W_3$, and $W_4$ bands.

On the other hand, we used photometric data from the Dark Energy Survey \citep[DES,][]{DESCollaboration+16}. This is a comprehensive photometric survey conducted with the 4-meter Victor M. Blanco Telescope at Cerro Tololo Inter-American Observatory. The survey has mapped approximately 5,000 square degrees of the southern Galactic cap in five broad optical bands: $g$, $r$, $i$, $z$, and $Y$. The survey achieves a $10\sigma$ depth of $i \sim 24$ (AB magnitude) for point sources, with a median full width at half-maximum of the point-spread function of 0.88 arcseconds in the $i$ band.

In this study, we used the photometric redshifts and galaxy classifications provided by \citet{Wen&Han22}, who employed the \texttt{SPREAD\_MODEL} parameter to distinguish galaxies from stars and calculated photometric redshifts using magnitudes from DES ($grizY$) and unWISE ($W_1$ and $W_2$). For consistency, we use the same photometric data to perform spectral energy distribution (SED) fitting and derive physical properties (Section \ref{sec:physical-parameters}). However, for all the other purposes (e.g., red sequence fits and magnitude cuts), we used the LS DR10 photometric data. A comparison between the magnitudes from the two surveys (DES and LS), and their impact on derived parameters, is presented in Appendix \ref{sec:appendix-comparison-between-surveys}.

\section{Galaxy properties}
\label{sec:galaxy-properties}

\subsection{Background and software}
\label{sec:background}

We utilize four parameters obtained through nonparametric methods: concentration index ($C$), asymmetry ($A$), Gini coefficient ($G$) and the moment of light $(M_{20})$. The mathematical formulation of each of them are described below (Section \ref{sec:non-parametric-approach}). To calculate them, we use the Python package \textsc{statmorph}\footnote{\url{https://statmorph.readthedocs.io/}} \citep{Rodriguez-Gomez19}. This code allows for the computation of several morphological parameters (e.g., $CAS$ statistics, \citealt{Conselice03}; $G$-$M_{20}$ statistics, \citealt{Lotz04}; $MID$ statistics, \citealt{Freeman+13}). As input, it requires providing a cutout of the science image, a segmentation map indicating which pixels belong to the galaxy, and the CCD gain with which the image was generated or the weight map associated with the observation. Additionally, an optional mask can be added to reduce contamination from nearby objects or the contributions of flux from bad pixels. In our analysis, we used masks to remove foreground stars, background galaxies, and image artifacts from the postage stamps before running \textsc{statmorph}. This step is crucial to avoid contamination in the calculation of structural parameters, particularly those that depend on the total light distribution such as the Petrosian radius, Gini coefficient, or asymmetry index. As discussed in \citet{Lotz04}, unmasked sources can bias these measurements, and the \textsc{statmorph} documentation\footnote{\url{https://statmorph.readthedocs.io/en/latest/}} similarly notes that unmasked neighbors may lead to discontinuities in the segmentation map and incorrect parameter estimation.

It should be noted that, with respect to the $CAS$ statistics, we only use the parameters $C$ and $A$. This choice is made because the resolution and signal-to-noise ratio (S/N) in our data are not sufficient to obtain reliable results for the smoothness parameter ($S$).

The concentration index is a parameter that has been widely used in classification systems. It is defined as the ratio between an outer and an inner radii that encloses a particular physical quantity. In the area of automated galaxy classification, this quantity is a fraction of the total flux. It is important to note that, when a radius that encloses a fraction of the total flux is defined, we are actually establishing a limit to the galaxy. It has been shown in the literature \citep[e.g., ][]{Conselice03,Lotz04} that using $1.5R_\text{P}$ as the ``total radius'' of the galaxy, where $R_\text{P}$ is the Petrosian radius, is representative for galaxies at different redshifts and allows comparison between them, since this radius is based on a curve of growth, which makes it largely insensitive to variations in the limiting surface brightness and S/N of observations \citep[][]{Lotz04}. The last important consideration in calculating this parameter is the choice of the center. For homogeneity, the center of asymmetry (see below) is used in the concentration and asymmetry calculations by \textsc{statmorph}. 

The asymmetry index allows us to quantify asymmetries in terms of the shape and flux distribution of the galaxy. To calculate the value of this parameter, the image of the galaxy must be rotated by a specific angle. Although different angles of rotation have been proposed in the literature, the most widely used is the 180° angle. Additionally, it is imperative to find a proper center of asymmetry to perform these calculations, which do not necessarily coincide with the centroid of the galaxy. Shifting this center by a few pixels can greatly alter the obtained value \citep{Conselice00}. For this task, the downhill simplex algorithm \citep{Nelder+Mead1965} is used to find the center that minimizes the asymmetry of the galaxy in a region close to its centroid.

The Gini coefficient is widely used in economy, as it shows the rank-ordered cumulative distribution function of a population's wealth. This statistic is based on the Lorenz curve, defined as

\begin{equation}
    L(p) = \dfrac{1}{\Bar{X}} \int_{0}^{p} F^{-1}(u) \,du,
\end{equation}

where $p$ is the fraction of the poorest citizens, $F(x)$ is the cumulative distribution function, and $\Bar{X}$ is the mean over all $X_i$ values \citep[][]{Lorenz05}.

Naturally, this definition was adapted in such a way that $G$ is a parameter to estimate the morphology of a galaxy, considering the flux (wealth) distribution in pixels that corresponds to the galaxy (population), rather than its application in economics \citep[][]{Abraham03,Lotz04}.

The second-order moment of the brightest 20\% region of a galaxy is the last nonparametric index used in this paper. It measures how far the pixels contributing to the brightest 20\% of the galaxy's light are located from its center.

It is important to note that both in the calculation of $G$ and in that of $M_{20}$, it is necessary to identify the pixels that belong to the galaxy, and in fact, the values of these parameters are very sensitive to this detection. A segmentation map is used to perform this task, which has been defined in at least two different ways. For example, \cite{Abraham03} used a constant surface brightness threshold to determine which pixels are part of the galaxy, i.e., pixels that lie above that threshold. On the other hand, \citet{Lotz04} proposed a method in which the ``Gini'' segmentation map depends only on the Petrosian radius, becoming insensitive to the surface brightness dimming of distant galaxies and allowing better comparison. Here, we lean toward the second option, supported by the advantage that this method allows for a better comparison between galaxies at different redshifts, and it is also the default method used in \textsc{statmorph}.

Specifically, to create the ``Gini'' segmentation map the galaxy image is first convolved with a Gaussian with $\sigma = R_P/5$. This step arises the signal of the galaxy pixels above the background noise, making low surface brightness galaxy pixels more detectable. Next, the surface brightness $\mu$ at $R_P$ is measured and pixels in the smoothed image with flux values greater than $\mu (R_P)$ and less than $10 \sigma$ from their neighboring pixels are assigned to the galaxy. This last step ensures that any remaining contamination (e.g., cosmic rays or spurious noise pixels) in the image are not included in the ``Gini'' segmentation map \citep{Lotz04}.
\\

\begin{figure}[h]
    \centering
    \includegraphics[width=1\linewidth]{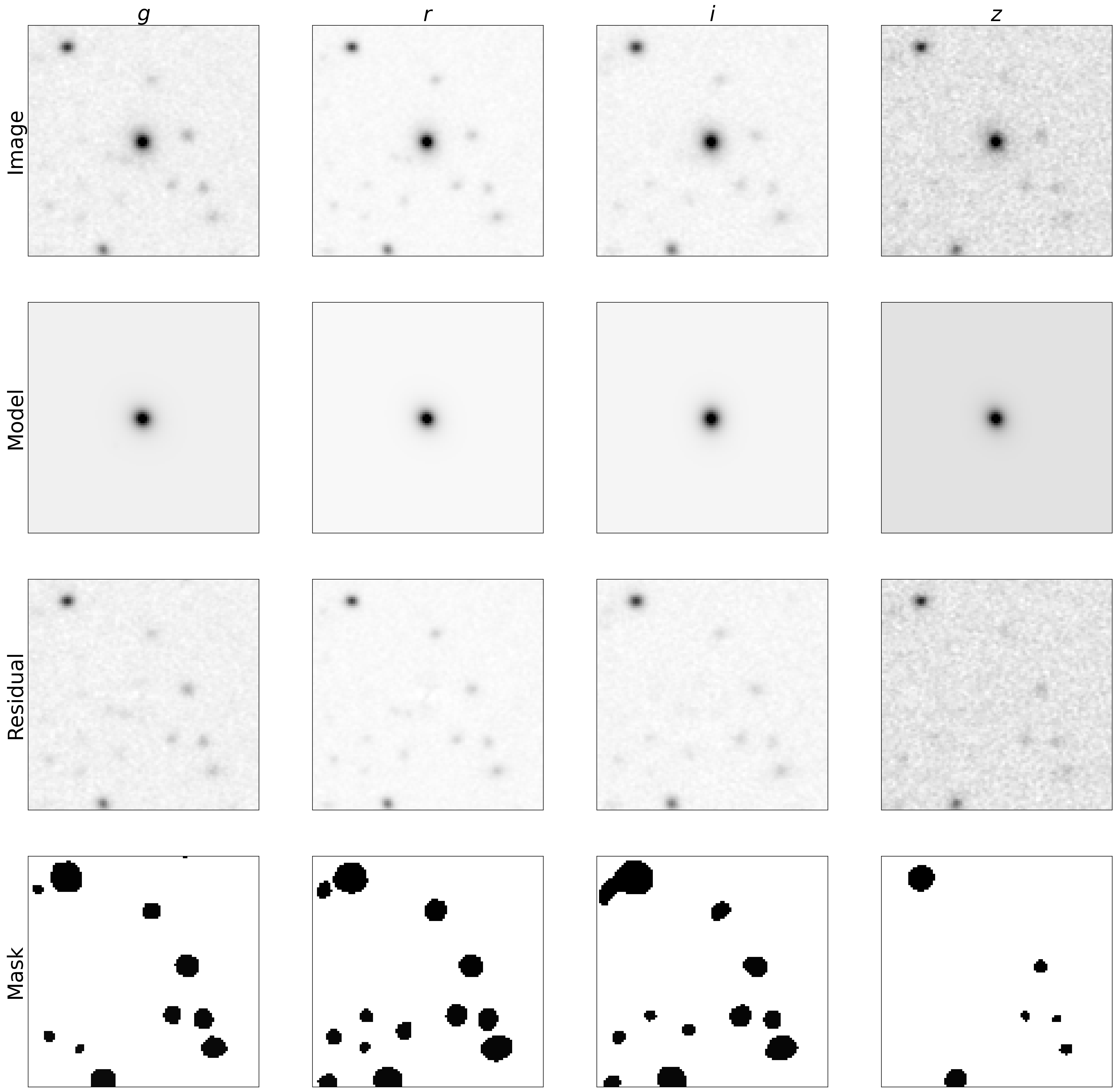}
    \caption{Example of the \textsc{Galfitm} output for the bulge-dominated galaxy J002712.03-343901.38 belonging to the Abell 2715 galaxy cluster, with a S\'ersic index value of $n_r = 2.527 \pm 0.079$. The residual image corresponds to the difference between the observed image and the fitted model.}
    \label{fig:galfitm-bulge-example}
\end{figure}

\begin{figure}[h]
    \centering
    \includegraphics[width=1\linewidth]{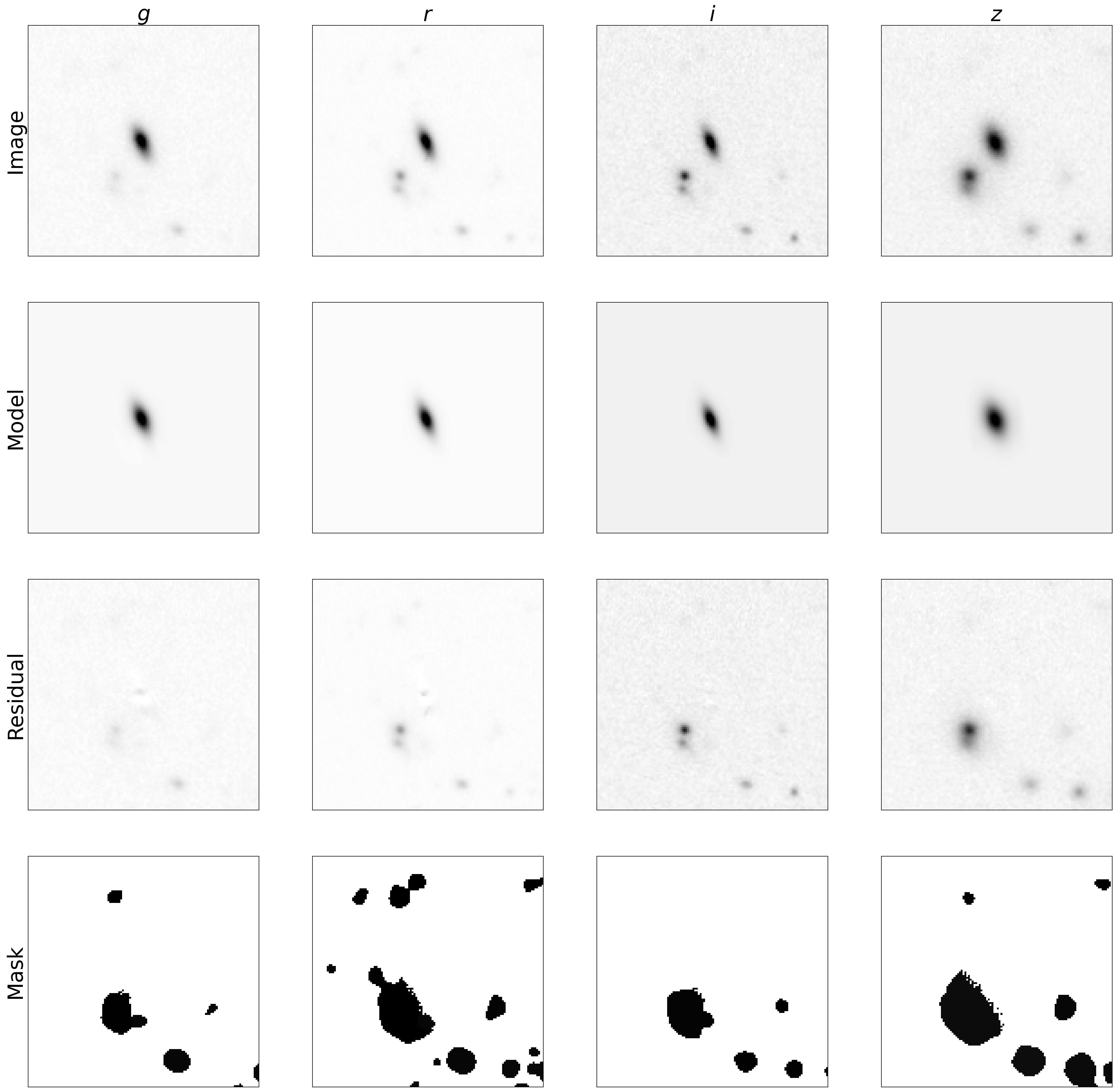}
    \caption{Example of the \textsc{Galfitm} output for the disk-dominated galaxy J220241.77-600131.8, belonging to the Abell 3827 galaxy cluster, with a S\'ersic index value of $n_r = 0.984 \pm 0.010$.}
    \label{fig:galfitm-disk-example}
\end{figure}

On the side of parametric methods for determining the morphology of galaxies, we derived the S\'ersic index \citep{Sersic68} and the effective radius using the \textsc{Galfitm} software from the \textsc{MegaMorph} project\footnote{\url{https://www.nottingham.ac.uk/astronomy/megamorph/}} \citep{Vika+13,Vika+15}. This code is an adaptation of \textsc{Galfit} \citep{Peng+02,Peng+10}, which enables fitting two-dimensional surface brightness profiles by minimizing the $\chi^2$ statistic. The advantage of \textsc{Galfitm} is that it allows simultaneous fits in images of different filters of the same galaxy, thus constraining these wavelength-dependent models. We fit single S\'ersic models using the four photometric bands available in the LS DR10 simultaneously. Each parameter involved in the fitting -- position, magnitude, effective radius, S\'ersic index, axis ratio, position angle, and sky background -- was allowed to vary with four degrees of freedom across the four photometric bands. In Figs. \ref{fig:galfitm-bulge-example} and \ref{fig:galfitm-disk-example} we show examples of the \textsc{Galfitm} outputs and the masks that we used for a bulge-dominated and a disk-dominated galaxy, respectively (see Section \ref{sec:morphological-classification} for details on the morphological type definitions).

The S\'ersic index has been used in the literature to distinguish between spheroidal and disk-type galaxies \citep[e.g.,][]{Buitrago+13, Vika+15}, or in conjunction with the effective radius to identify populations in different environments \citep[e.g.,][]{Montaguth+23}.

\subsection{Nonparametric approach}
\label{sec:non-parametric-approach}

The mathematical expression of the concentration index used by \textsc{statmorph} follows the definition of \citet{Conselice03}:

\begin{equation}
    C = 5 \, \log\left(\dfrac{r_{80}}{r_{20}}\right),
\end{equation}

where $r_{80}$ and $r_{20}$ are the radii enclosing 80\% and 20\% of the galaxy light, respectively.

The asymmetry is calculated as follows \citep[][]{Abraham96,Brinchmann98,Conselice03}:

\begin{equation}
    A = \dfrac{\Sigma_{i,j} |I_{ij} - I_{ij}^{180}|}{\Sigma_{i,j} |I_{ij}|} - A_{\text{bg}},
\end{equation}

where $I_{ij}$ and $I_{ij}^{180}$ are the pixel flux values within $1.5R_{\text{P}}$ of the original and rotated images, respectively, and $A_{\text{bg}}$ is the asymmetry of the background, which is calculated over a square region of an area similar to that covered by the galaxy.

In the case of the Gini coefficient applied to morphological classification, the mathematical expression evolved until it reached the form described by \cite{Lotz04}:

\begin{equation}
    G = \dfrac{1}{|\Bar{X}| n (n-1)} \sum_{i}^{n} (2i - n - 1)|X_i|,
\end{equation}

where $n$ is the number of pixels belonging to a galaxy. In this context, $X_i$ corresponds to the pixel values.

For the second moment of light $M_{20}$, as introduced by \cite{Lotz04}, in order to measure it, first $M_{\text{tot}}$ must be found, which corresponds to the total second-order moment of the galaxy:

\begin{equation}
    M_{\text{tot}} = \sum_{i}^{n} M_i = \sum_{i}^{n} f_i \left[ (x_i - x_c)^2 + (y_i - y_c)^2 \right],
\end{equation}

where $f_i$ is the flux in each pixel, ($x_i$, $y_i$) the position of each pixel, and ($x_c$, $y_c$) is the center of the galaxy, calculated in such a way that $M_{\text{tot}}$ is minimized.

Then, it is necessary to rank-order the galaxy pixels by flux, and then:

\begin{equation}
    M_{20} \equiv \log \left( \dfrac{\Sigma_i M_i}{M_{\text{tot}}} \right) \text{, while} \sum_{i} f_i < 0.2 f_{\text{tot}}.
\end{equation}

In this equation, $f_{\text{tot}}$ is the total flux of pixels belonging to the galaxy, and $f_i$ are the fluxes for each pixel $i$, arranged in such a way that $f_1$ is the brightest pixel, $f_2$ is the second brightest pixel, and so on.

\subsection{Parametric approach}
\label{sec:parametric-approach}

The S\'ersic profile is a mathematical function that describes the intensity profile of a galaxy. This can be expressed using the following formulation:

\begin{equation}
    I(r) = I_\text{e} \exp \left\{ -b_n \left[ \left( \frac{r}{r_\text{e}} \right)^{(1/n)} - 1 \right] \right\},
\end{equation}

where $I$ is the intensity at position $r (x, y)$, $r$ is the radius from the center that corresponds to $(x, y)$, $r_e$ is the effective radius (half-light radius), $I_\text{e}$ is the intensity at the half-light radius ($I_\text{e} = I(r_\text{e})$), and $n$ is the Sérsic index which determines the slope of the profile. Values of $n=4$ correspond to a DEV profile, and $n=1$ indicates an exponential profile.

The constant, $b_n$, is defined so that $r_\text{e}$ contains half of the total flux and can be solved numerically using

\begin{equation}
    \Gamma(2n) = 2\gamma(2n, b_n),
\end{equation}

where $\Gamma$ and $\gamma$ are the complete and incomplete gamma functions, respectively.

Although the effective radius obtained from two-dimensional surface brightness S\'ersic models is a parameter commonly used for galaxy comparisons, the results can be misleading, as the galaxy radius is strongly dependent on its mass \citep{Sazonova+20}. Therefore, we also employed the definition of compactness, which is very similar to the surface mass density of galaxies:

\begin{equation}
    \Sigma = \dfrac{M_*}{\pi R_e^2},
\end{equation}

where $M_*$ is the stellar mass (whose calculation is detailed in Section \ref{subsubsec:stellar-mass-and-ssfr}) of the galaxy and $R_e$ is the effective radius obtained from S\'ersic models.

\subsection{Physical parameters}
\label{sec:physical-parameters}

\subsubsection{Normalized colors}

The colors of galaxies are a fundamental property. It has been shown that it is possible to clearly separate populations using this physical parameter, as it exhibits a notable bimodality (i.e., red galaxies and blue galaxies). However, because we are working with clusters located at different redshifts, it is necessary to apply corrections to compare populations across all systems. Essentially, we need to take into account the Doppler effect, whereby the light from galaxies is redshifted as they are moving away from us due to the expansion of the Universe.

To carry out this process, we subtracted the characteristic magnitudes ($m^*$) of the galaxy clusters in which they are located from the magnitudes of the galaxies. The characteristic magnitudes of these structures were calculated using \textsc{EzGal} \citep{Mancone&Gonzalez12}, modeling composite stellar populations (CSPs), as is explained in \citetalias{VelizAstudillo+2024}.

Thus, corrected (or normalized) colors $(g-r)_{\text{norm}}$ can be expressed as

\begin{equation}
    (g-r)_{\text{norm}} = (g - m^*_g) - (r - m^*_r), 
\end{equation}

where $m^*_g$ and $m^*_r$ are the characteristic magnitudes in the $g$ and $r$ bands, respectively.

\subsubsection{Stellar mass and specific star formation rate}
\label{subsubsec:stellar-mass-and-ssfr}

To determine the mass and specific star formation rate (sSFR) of galaxies, we used the \textsc{Le Phare}\footnote{\url{https://www.cfht.hawaii.edu/~arnouts/LEPHARE/lephare.html}} code \citep{Arnouts+99,Ilbert+06}. This Fortran code was designed to perform SED fitting to obtain the photometric redshift of galaxies, as well as some of their physical properties.

These calculations were the only ones conducted using photometric data from DES instead of LS DR10. This choice was driven by the availability of the $Y$ band in DES, allowing the use of seven filters for SED fitting. Specifically, we utilized the photometric information from DES, namely $grizY$, and from unWISE, which includes the $W_1$ and $W_2$ bands.

Regarding the SED libraries, we utilized a library of synthetic spectra generated with the stellar population synthesis model by \citet{BruzualCharlot03} with a Chabrier initial mass function (IMF). Three different metallicities were used (0.02, 0.04, and 0.008). We assumed an exponentially declining star formation history (SFH) with eight values for the characteristic timescale, $\tau$, ranging from 0.1 to 30 Gyr. The extinction curve of \citet{Calzetti+00} was considered, with 12 possible values for $E(B-V)$, ranging from 0.0 to 1.0 mag. In addition, the contribution of emission lines to SEDs was included, specifically modeling the lines of $[\text{O}{\text{II}}]$ ($\lambda$3727), $[\text{O}{\text{III}}]$ ($\lambda$4959, $\lambda$5007), H${\beta}$ ($\lambda$4861), H${\alpha}$ ($\lambda$6563), and Ly$_{\alpha}$ ($\lambda$1216). The parameters of the generated SED libraries are summarized in Table \ref{tab:sed-libraries-parameters}.

\begin{table}[h]
    \caption{Parameters used to create the SED libraries.}
    \label{tab:sed-libraries-parameters}
    \centering
    \resizebox{0.49\textwidth}{!}{
    \begin{tabular}{lc}
    \hline
    Parameter & Value(s)\\
    \hline
    Stellar population synthesis model & \citet{BruzualCharlot03}\\
    Initial mass function & Chabrier\\
    Metallicity $Z$ & 0.02, 0.04, 0.008\\
    Star formation history & Exponentially declining\\
    Characteristic duration $\tau$ [Gyr] & 0.1, 0.3, 1, 2, 3, 5, 10, 30\\
    Extinction curve & \citet{Calzetti+00}\\
    Color excess $E(B-V)$ [mag] & 0, 0.05, 0.1, 0.2, 0.3, 0.4, 0.5, 0.6, 0.7, 0.8, 0.9, 1\\
    Emission lines & $[\text{O}_{\text{II}}]$, $[\text{O}_{\text{III}}]$, H$_{\beta}$, H$_{\alpha}$, Ly$_{\alpha}$\\
    \hline
    \end{tabular}
    }
\end{table}

\subsection{Quality cuts}
\label{sec:quality-cuts}

To ensure the robustness of our results, we decided to apply several quality cuts:

\begin{enumerate}
    \item We applied the criterion $\sigma_{z_{\text{phot}}} < 0.05(1+z_{\text{phot}})$, thereby excluding galaxies with imprecise photometric redshifts, where $\sigma_{z_{\text{phot}}}$ represents the error of the photometric redshift.
    \item Following the \textsc{statmorph} recommendations, we decided to eliminate galaxies with a low S/N using the condition \texttt{sn\_per\_pixel}$(r) < 2.5$. Furthermore, this code provides a \texttt{flag} indicating the quality of the morphological measurements, taking values of 0 (good), 1 (suspicious), 2 (bad), and 4 (catastrophic). We applied the criterion \texttt{flag}$(r) \leq 1$.
    \item We chose a threshold for the $\chi^2_{\nu}$ parameter of \textsc{Galfitm} below which the galaxies are considered for the study. The maximum non-outlier value of $\chi^2_{\nu}$ corresponds to 0.89, but considering that $\chi^2_{\nu} = 1$ is the ideal fit, we used this last value as the threshold.
    \item Similarly to the third quality cut, we chose the maximum non-outlier value of the $\chi^2$ distribution obtained from \textsc{Le Phare} as the threshold below which the galaxies pass the filter, resulting in $\chi^2 = 26.56$.
    \item We retained only those galaxies within the range $-15 \leq \log(\text{sSFR}) \leq -8$ of the sSFR, following \cite{Chen+24}.
    \item We cut $\log(M_*/M_{\odot}) \geq 8.5$ to minimize the issue of mass completeness \citep{Chen+19,Liu+19}.
    \item We only worked with galaxies with a magnitude $m < m^* + 3$ in the $r$ band, where $m^*$ is the characteristic magnitude of each cluster. This ensures a homogeneous study in terms of photometric depth and also ensures the reliability of the LS DR10 photometric measurements.
\end{enumerate}

We note that all brightest cluster galaxies were also excluded at this point. Their distinct evolutionary paths, dominated by late-time mergers and accretion processes, as well as potential star formation fueled by cooling flows in the cluster core, result in structural and physical properties that deviate from those of typical cluster member galaxies \citep[e.g.,][]{Lin+Mohr2004,Rawle+2012,Haines+2013}. At this stage, our sample comprised 8,412 galaxies distributed among 87 clusters, with an average number of reliable cluster members of $N_{\text{mem}} = 97$ after applying quality cuts.

\section{Morphological classification}
\label{sec:morphological-classification}

To perform morphological classification, we followed the approach of \citet{Sazonova+20}, whereby they used principal component analysis (PCA) to scale the dataset ${G, M_{20}}$ and then classify galaxies. Principal component analysis serves as a statistical method that utilizes linear transformation to streamline the dimensionality of a dataset. This is achieved by identifying variables that are linearly correlated, thereby reducing noise and redundancy within the data. One key benefit of PCA is its ability to convert a collection of potentially interrelated variables into a new set of more fundamental, independent variables \citep{Hotelling33}. Furthermore, when there is significant redundancy and, consequently, correlations among the variables, PCA can often reconstruct the original variable values using fewer principal components than were present in the original dataset, preserving essential features.

The main goal of applying PCA is to normalize the data and find the main sequence that divides normal Hubble-type galaxies and mergers, as well as the separation between early-type and late-type galaxies. When PCA is applied to only two variables, the first component corresponds to the main sequence of the data, and the second component corresponds to the standard deviation of this sequence. This method offers the benefit that once there are sufficiently reliable measurements for the morphological parameters, spatial resolution no longer plays a crucial role in the classification system since the dataset for ${G, M_{20}}$ is standardized.

\begin{figure*}[h]
    \centering
    \includegraphics[width=1\textwidth]{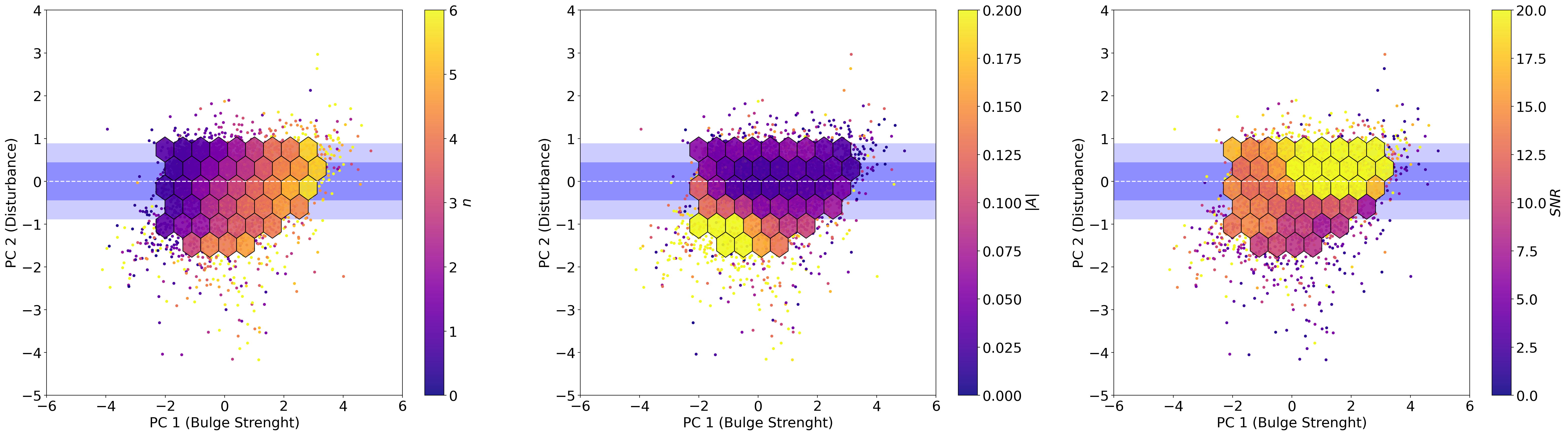}
    \caption{Principal components (PC 1 and PC 2) of the morphological parameters $G$ and $M_{20}$ for approximately 17,000 galaxies. A hexagonal binning has been applied with a grid size of (30, 9), considering only bins containing at least 20 data points. The horizontal dashed white line corresponds to the main sequence in this plane, and the shaded regions in dark and light blue correspond to the 1$\sigma$ and 2$\sigma$ zones, respectively. The hexbins follow a color map associated with three different variables. From left to right, we have: the Sérsic index, $n$, asymmetry, $A$, and \textit{S/N}.}
    \label{fig:pca-morphology-interpretation}
\end{figure*}

In Fig. \ref{fig:pca-morphology-interpretation}, we have the results of the PCA. In the left panel, we notice that within the 2$\sigma$ confidence interval, there is a clear trend of galaxies to increase the value of PC 1 along with the S\'ersic index, $n$. For this reason, we interpret the first principal component as the bulge strength, indicated in the X-axis label. Then, observing the middle panel, we see that within this same confidence interval, galaxies have an absolute value of asymmetry very close to zero. This leads us to interpret the second principal component as the degree of disturbance that galaxies exhibit. However, as we can see in the right panel, there is also a trend for galaxies below the 2$\sigma$ confidence zone to have a lower S/N. Therefore, considering this, we can conclude that a galaxy with a very positive value of PC 2 implies that it is an asymmetric galaxy and then a merger candidate. But if the value of PC 2 is very negative, then it is a diffuse or unresolved galaxy \citep{Sazonova+20}.

However, terms like ``very high PC 2'' or ``very low PC 2'' are not precise at all. Although there is a trend for bulge strength to increase with PC 1, it is pertinent to define the classification system clearly. Galaxies with $\text{PC 2} > 1$ are considered candidates for mergers, and those with $\text{PC 2} < -2$ are classified as diffuse/unresolved galaxies. Galaxies with $-2 \leq \text{PC 2} \leq 1$ are considered normal Hubble-type galaxies. Those normal Hubble-type galaxies with $\text{PC 1} \leq -0.5$ are disk-dominated and those with $\text{PC 1} > -0.5$ are bulge-dominated. The reason for choosing the threshold $\text{PC 1} = -0.5$ to separate bulge and disk galaxies, instead of the value of $\text{PC 1} = 1$ used by \citet{Sazonova+20}, is because in our data, this bin has a mean value for the S\'ersic index of $n \sim 2.5$, a value commonly used in the literature to separate early-type and late-type galaxies \citep[e.g.,][]{Buitrago+13}. Moreover, to be more strict in the classification and following this S\'ersic index criteria, we also imposed the condition that bulge-dominated and disk-dominated galaxies must have $n > 2.5$ and $n < 2.5$, respectively. In total, our morphological classification yields 2,517 bulge-dominated and 2,515 disk-dominated galaxies.

\section{Results}
\label{sec:results}

Before comparing the properties of galaxies in relaxed and disturbed clusters, we performed a mass-match between both subsamples. This is necessary because stellar mass is a key factor driving galaxy evolution. We used a tolerance of 0.05\%, obtaining 2,505 galaxies in each cluster type (relaxed, intermediate, and disturbed). However, as our objective is to study the populations between the most relaxed and most disturbed clusters, we excluded the sample of intermediate systems. Thus, for the analysis, we have 5,010 galaxies, split equally between relaxed and disturbed clusters, with statistically similar mass distributions.

\subsection{Mass-size relations}
\label{sec:mass-size-relations}

We studied the mass-size relation by classifying galaxies in two different ways. On the one hand, we separated galaxies by morphological types (Section \ref{sec:morphological-classification}). On the other hand, we separated galaxies according to their quenching state. For this, we employed the criteria whereby quenched galaxies are those with $\log(\text{sSFR}/[\text{yr}^{-1}]) < -11$, and star-forming galaxies are those with $\log(\text{sSFR}/[\text{yr}^{-1}]) \geq -11$ \citep{Wetzel+12}. 

\begin{figure}[h]
    \centering
    \includegraphics[width=1\linewidth]{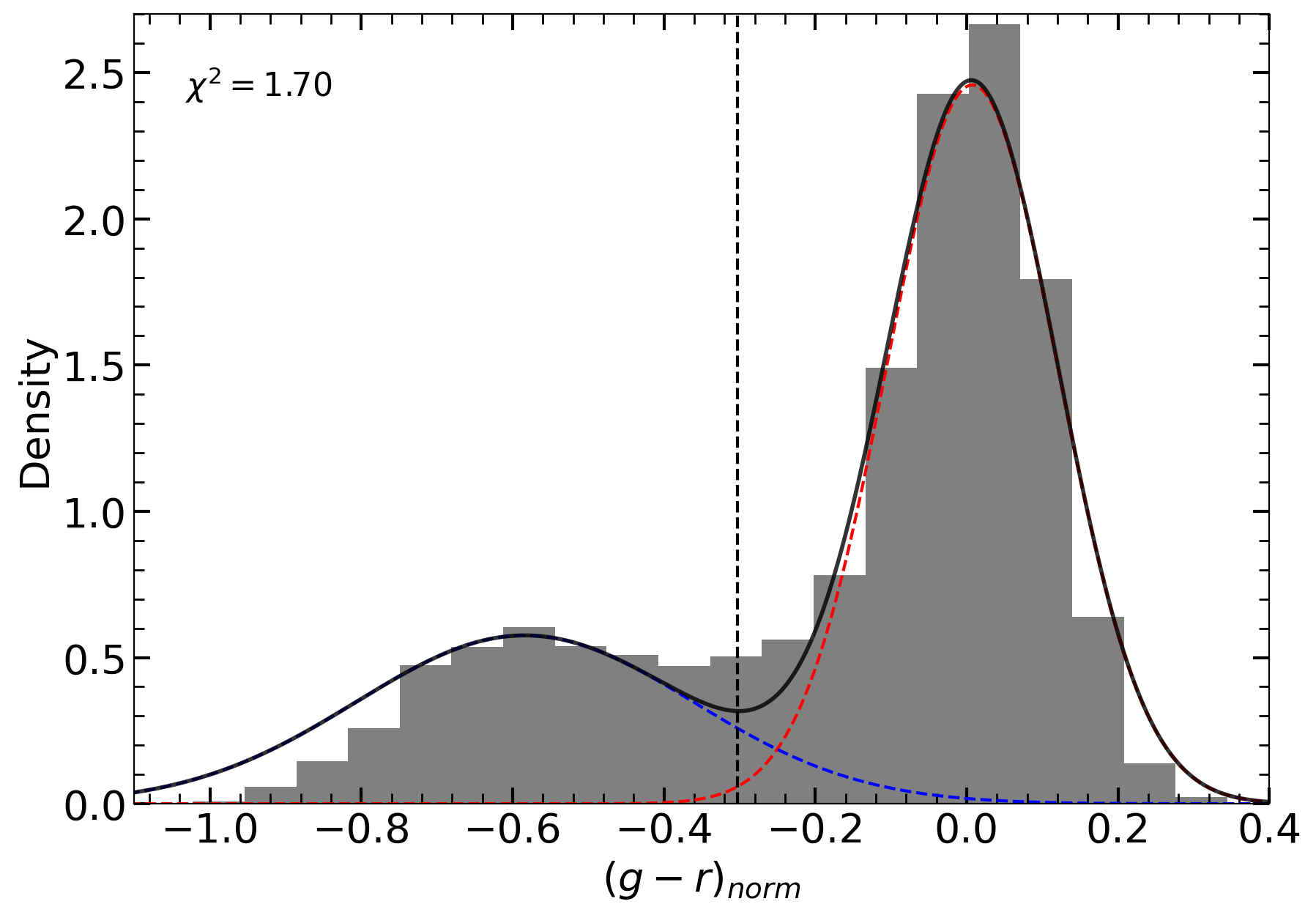}
    \caption{Color distribution of all galaxies after applying the quality cuts (Section \ref{sec:quality-cuts}). Dashed blue and red curves correspond to the Gaussian models fitted to the blue and red component, respectively, while the solid black curve is the sum of both. The vertical dashed black line is the color threshold to separate star-forming blue and passive red galaxies, selected as the minimum value between the two peaks, $(g-r)_{\text{norm}} = -0.3$.}
    \label{fig:color-threshold}
\end{figure}

However, considering the uncertainties in the determination of the stellar mass and the SFR, using only the sSFR criterion may not be sufficient \citep{Chen+24}. To address this, we also applied a color cut based on the bimodality of the color distribution. As is seen in Fig. \ref{fig:color-threshold}, we fit a double Gaussian model using the Python package \textsc{lmfit}, and we used the minimum value between the two peaks as the color threshold to separate red and blue galaxies, i.e., $(g-r)_{\text{norm}} = -0.3$. In this way, we classified galaxies as star-forming or quiescent if they matched the SFR and color conditions, obtaining 1,922 quenched galaxies, 1,213 star-forming galaxies, and 1,875 unclassified galaxies (i.e., ones that do not satisfy the criteria of sSFR and color simultaneously).

\begin{figure}[h]
    \centering
    \includegraphics[width=1\linewidth]{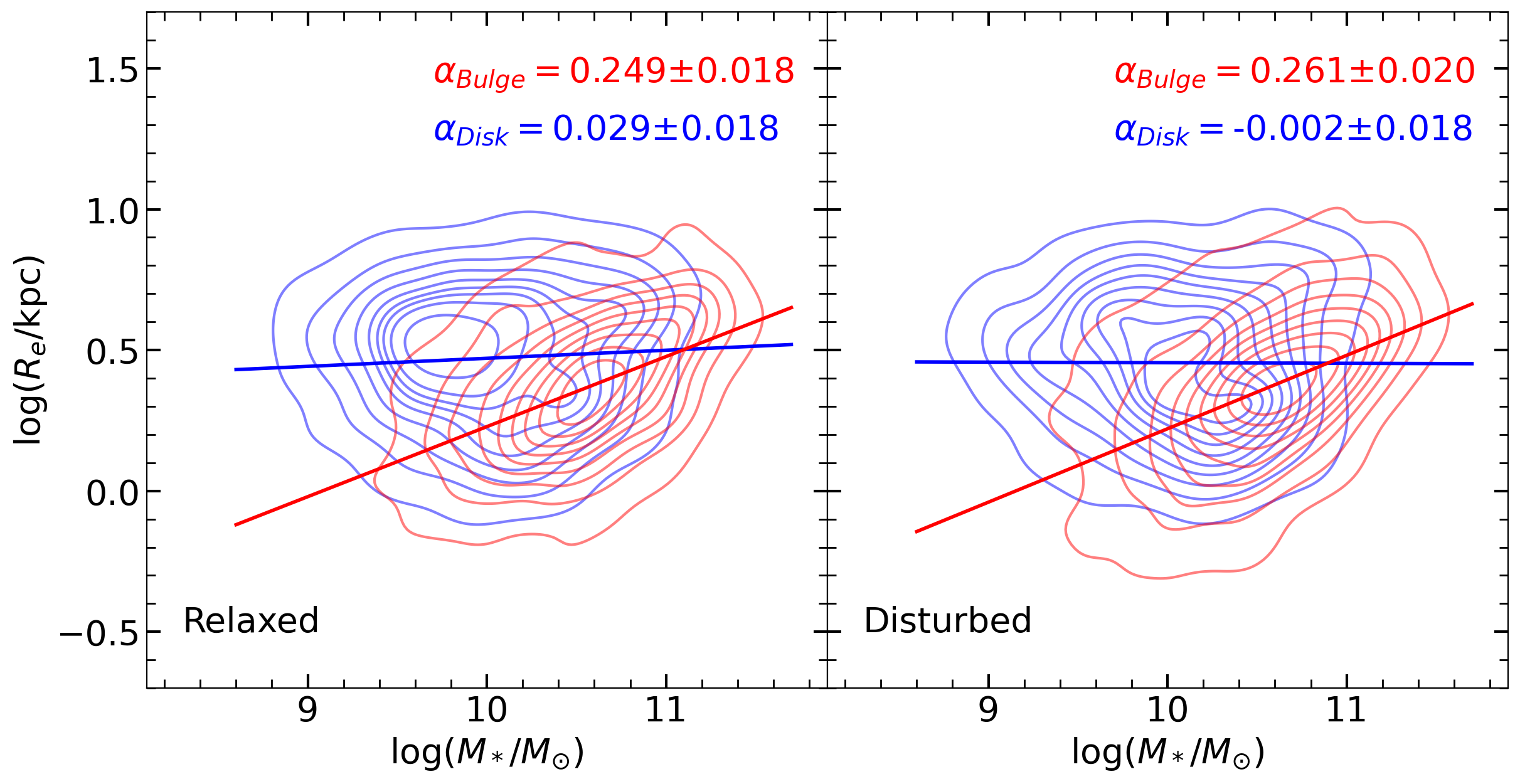}
    \caption{Distribution of bulge galaxies (red contours) and disk galaxies (blue contours) in the mass-size logarithmic space, separated into relaxed and disturbed clusters. The solid lines show the linear regressions applied to the data (following the same color code), and the text indicates the slopes with their respective errors for each galaxy population.}
    \label{fig:mass-size-relation-morphology}
\end{figure}

\begin{figure}[h]
    \centering
    \includegraphics[width=1\linewidth]{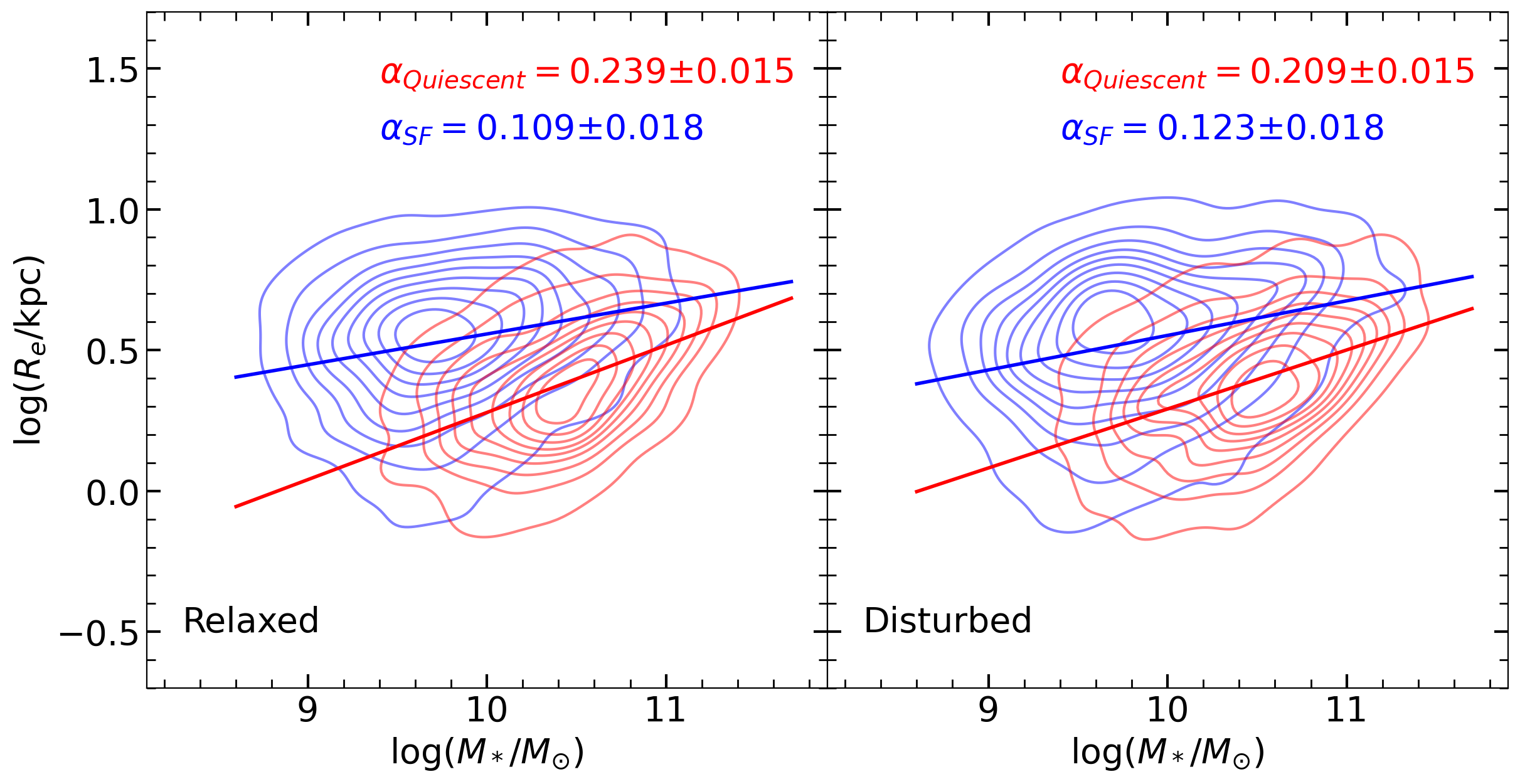}
    \caption{Similar to Fig. \ref{fig:mass-size-relation-morphology}, but separating galaxies into quiescent and star-forming.}
    \label{fig:mass-size-relation-sfr}
\end{figure}

In Figs. \ref{fig:mass-size-relation-morphology} and \ref{fig:mass-size-relation-sfr}, we depict the mass-size relations for morphology and star formation state. We also show the slopes of the linear regressions fitted with \textsc{lmfit} and their respective errors.

Although we visually observe some differences in the mass-size relations between relaxed and disturbed clusters, it is essential to quantify these differences statistically. To achieve this, we utilized the bootstrap resampling method. This statistical technique involves repeatedly resampling our dataset with replacement to generate a distribution of the measured parameters, enabling us to estimate robust confidence intervals for these parameters without relying on assumptions of normality. We adopted a 95\% confidence level, corresponding to a significance threshold of approximately $2\sigma$ for normally distributed data. Thus, differences in parameters between relaxed and disturbed clusters are considered statistically significant if the bootstrap confidence intervals at this level do not overlap. In other words, statistical significance is confirmed if zero is not contained within the confidence interval of the difference.

\begin{table}[h]
    \caption{Results of the bootstrap analysis comparing slopes of linear regressions for relaxed and disturbed clusters in the mass-size logarithmic space.}
    \label{tab:p-values-mass-size}
    \centering
    \resizebox{0.3\textwidth}{!}{
    \begin{tabular}{lr}
    \hline
    Galaxy type & Confidence interval \\
    \hline
    Morphology \\
    Bulge & [-0.013, 3.247] \\
    Disk & [-0.155, 0.723] \\
    \hline
    Star-formation rate \\
    Quiescent & [-0.585, 0.571] \\
    Star-forming & [-1.227, 0.117] \\
    \hline
    \end{tabular}
    }
\end{table}

\subsection{Morphology and SFR versus local environment}
\label{sec:morphology-sfr-vs-local-environment}

To investigate whether there are differences in the relationship between galaxies and their local environment, we used the normalized clustercentric distance ($R/R_{200}$) and the local density. The latter parameter is defined as $\Sigma_{10} = 10/A$, where $A$ is the circular area containing the ten nearest neighbors with $M_*/M_{\odot} \geq 10.5$ of a galaxy \citep{Dressler80,Vulcani+23b}.

\begin{figure}[h]
    \centering
    \includegraphics[width=1\linewidth]{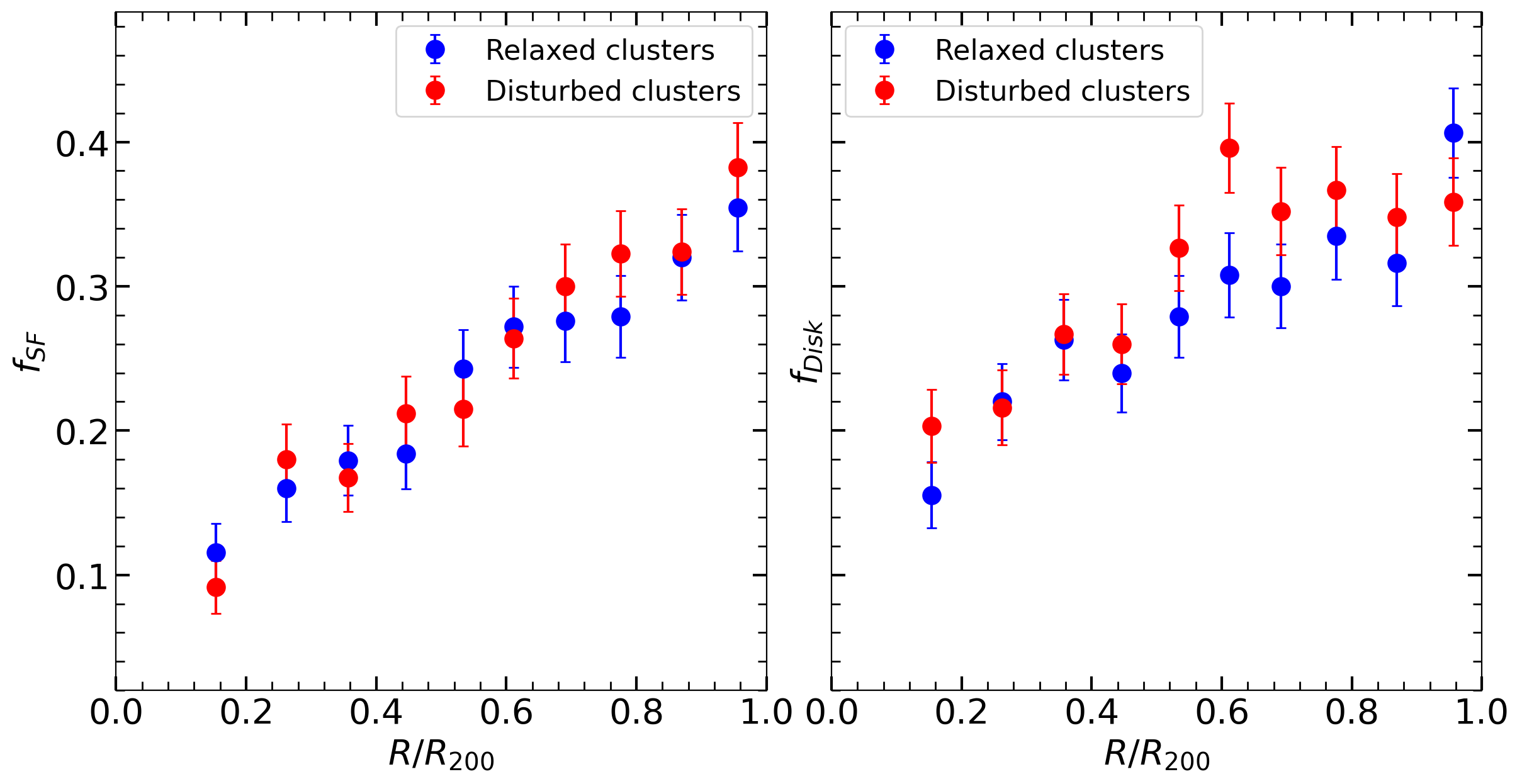}
    \caption{SF- and morphology-clustercentric distance relation for relaxed and disturbed clusters (from left to right). They are 250 galaxies in each bin. The color points show the fraction of star-forming galaxies (left panel) and disk-dominated galaxies (right panel), while the error bars indicate the uncertainties estimated using Poisson errors. Blue and red symbols correspond to relaxed and disturbed clusters, respectively.}
    \label{fig:morph-sfr-clustercentric-distance}
\end{figure}

\begin{figure}[h]
    \centering
    \includegraphics[width=1\linewidth]{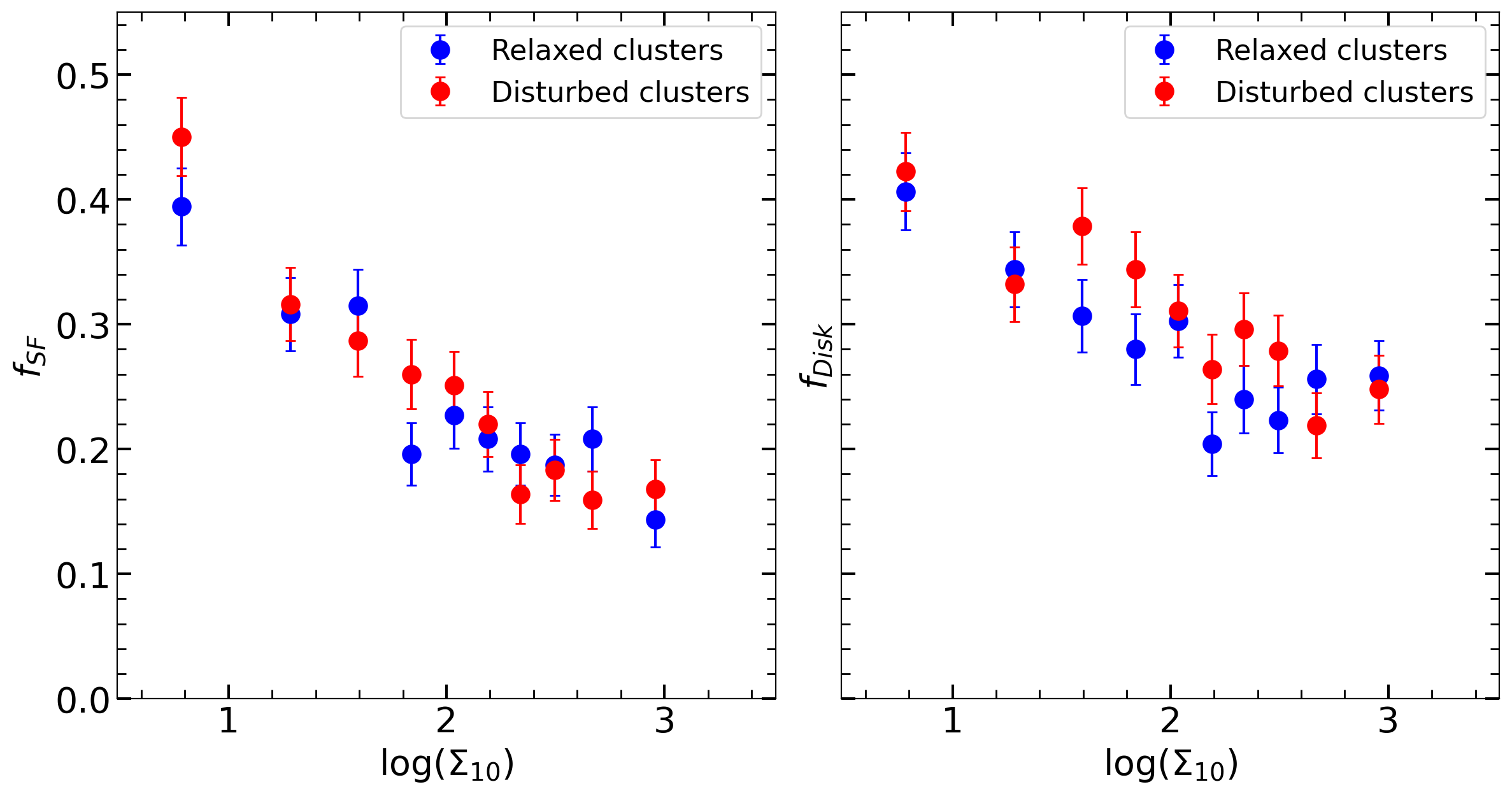}
    \caption{Similar to Fig \ref{fig:morph-sfr-clustercentric-distance}, but for the morphology- and SF-density relation.}
    \label{fig:morph-sfr-local-density}
\end{figure}

In Fig. \ref{fig:morph-sfr-clustercentric-distance} we display the fractions of galaxy types according to star formation state (left panel) and morphology (right panel) as a function of the normalized clustercentic distance, whereas in Fig. \ref{fig:morph-sfr-local-density} we present the same fractions but as function of the local density. We note that the we use the position of the X-ray peak as the center of the galaxy clusters. Each clustercentric distance and local density bin contains 250 galaxies.

For the morphology-clustercentric distance and SF-clustercentric distance relations, we observe that the fraction of disk-dominated and star-forming galaxies increases at larger distances for both relaxed and disturbed clusters. Qualitatively, we do not find significant differences in the trends as a function of the dynamical state. In the case of the morphology-density and SF-density relations, the general trend is a decrease in the fraction of star-forming and disk-dominated galaxies with increasing local density, with a slightly higher fraction of disk-dominated galaxies in disturbed clusters in two specific bins.

\subsection{Color-magnitude diagram}
\label{sec:color-magnitude-diagram}

\begin{figure}[h]
    \centering
    \includegraphics[width=1\linewidth]{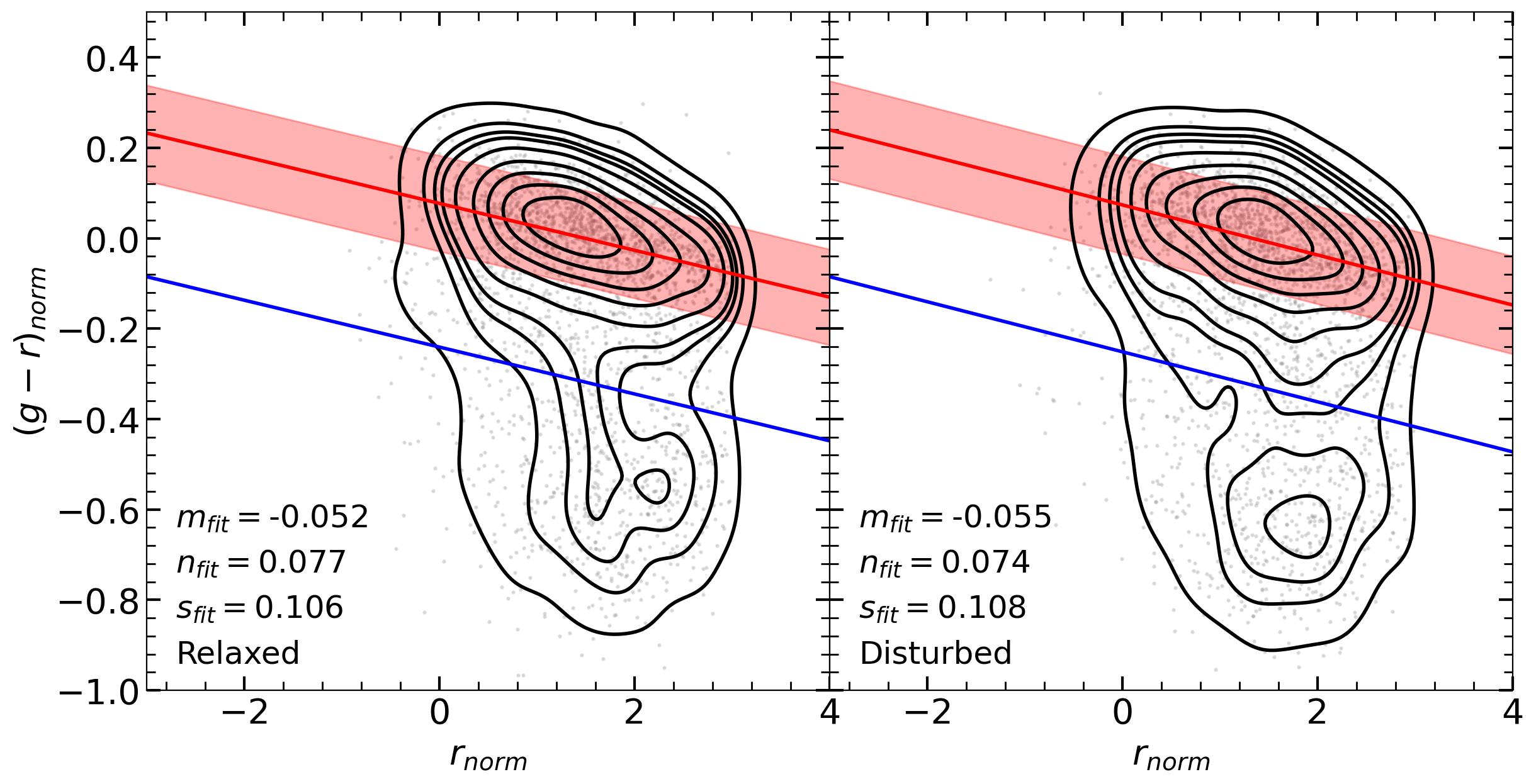}
    \caption{Normalized and stacked CMDs, separating the sample into relaxed and disturbed clusters. Each gray point corresponds to a galaxy. The solid red line represents the best fit of the red sequence, while the shaded red area corresponds to the $\pm 1 \sigma$ region. The solid blue line is the same best fit of the red sequence line but shifted $3\sigma$ downward. The text shows the parameters of the robust linear regressions.}
    \label{fig:color-magnitude-diagrams}
\end{figure}

In both subsamples (i.e., relaxed and disturbed clusters), we employed the same robust linear regression model as in Paper I, i.e. Huber regressor, for individual red cluster sequence (RCS) fits. The normalized color-magnitude diagrams (CMDs) for each environment are shown in Fig. \ref{fig:color-magnitude-diagrams}, along with the regression parameters. We calculated the error of these parameters using 10,000 bootstrap iterations and robust biweight estimators, but considering that all of them are on the order of $10^{-5}$ we decided not to include them in the text of Fig. \ref{fig:color-magnitude-diagrams}.

In this context, we applied the bootstrap method to assess whether significant differences exist between the parameters of the RCSs derived for relaxed and disturbed clusters. The results of this analysis are presented in Table \ref{tab:p-values-color-magnitude}. We do not find statistically significant differences between relaxed and disturbed cluster parameters, as all 95\% confidence intervals of the parameter differences include zero.

\begin{table}[h]
    \caption{Results of the bootstrap analysis applied to the parameters of robust linear regressions performed on red galaxies to derive the red sequences in relaxed and disturbed galaxy clusters.}
    \label{tab:p-values-color-magnitude}
    \centering
    \resizebox{0.28\textwidth}{!}{
    \begin{tabular}{lr}
    \hline
    Parameter & Confidence interval \\
    \hline
    Slope & [-1.841, 0.012] \\
    Intercept & [-0.010, 3.416] \\
    Scatter & [0.000, 1.015] \\
    \hline
    \end{tabular}
    }
\end{table}

On the other hand, we can use the CMDs to delineate subpopulations within each cluster type. We define red sequence galaxies as those falling within $\pm 1 \sigma$ of the best fit, while blue cloud (BC) galaxies are identified as those lying beyond $3\sigma$ from the best fit. Meanwhile, we define the green valley (GV) galaxies as those which lie in the intermediate zone between these two populations.

\subsection{Parameters comparison}
\label{sec:parameters-comparison}

In addition to the populations defined in the previous sections regarding morphology and color, we now present a new division. This time, we separated high-mass galaxies ($\log(M_*/M_{\odot}) > 10.5$) from low-mass galaxies ($\log(M_*/M_{\odot}) \leq 10.5$).

With the populations defined, we compared the distributions of the physical and structural parameters of galaxies in relaxed and disturbed clusters. For this, we used the Kolmogorov-Smirnov (KS) test with a confidence level of 99\%, meaning that distributions with a $p$ value of $< 0.01$ are considered statistically different.

\begin{figure*}[h]
    \centering
    \includegraphics[width=1\linewidth]{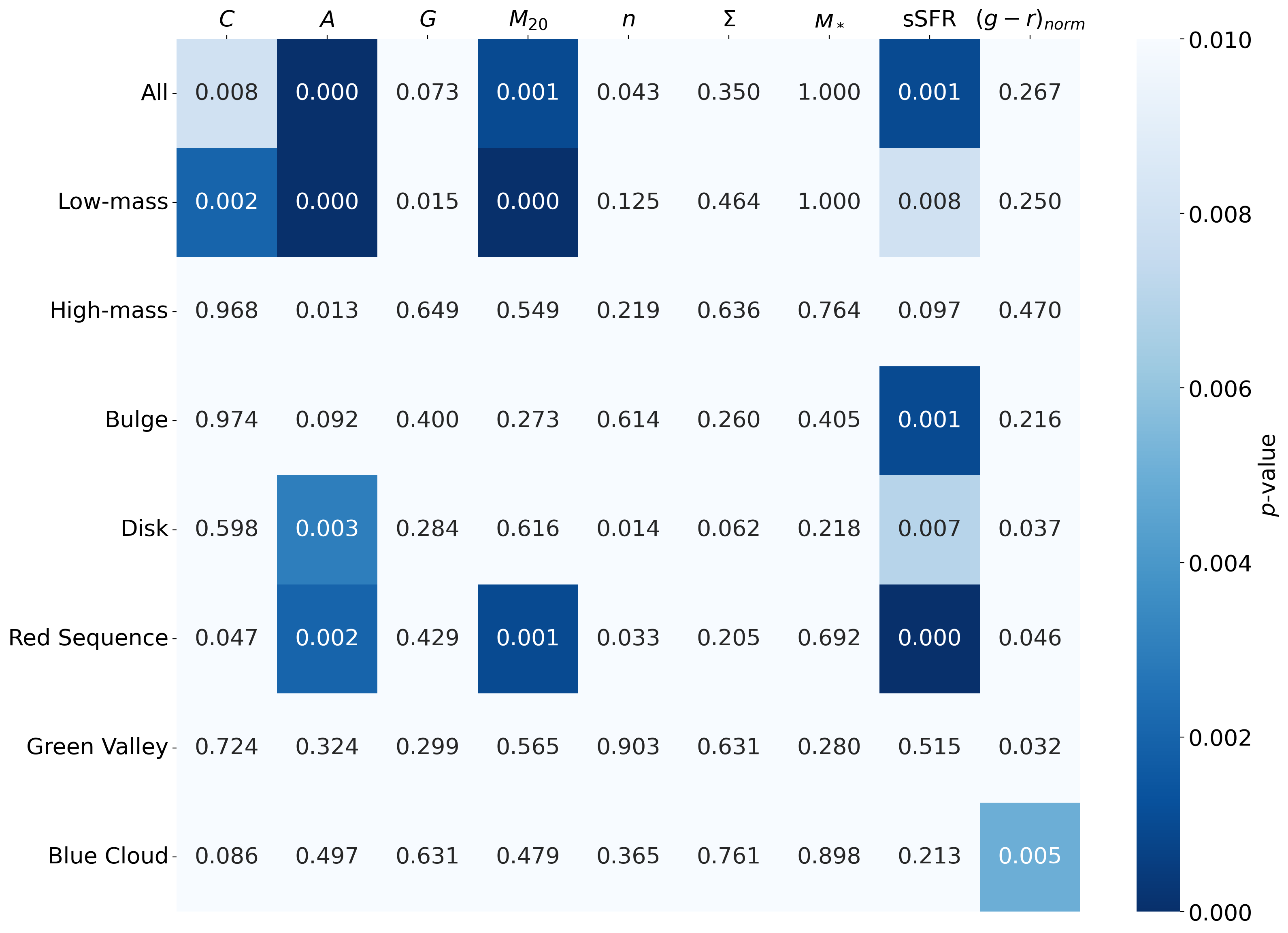}
    \caption{Summary of the results of applying the KS test to all the cluster populations studied in this work. This heatmap presents the $p$ values that result from comparing relaxed with disturbed clusters.}
    \label{fig:parameters-distribution-comparison}
\end{figure*}

In Fig. \ref{fig:parameters-distribution-comparison}, we present a summary of the results from the KS tests, highlighting in a more intense blue color those distributions that show the greatest differences between relaxed and disturbed clusters. We note that the dynamical state of the clusters has a differential effect on their member galaxies, with low-mass and red sequence galaxies being the most affected.

\begin{table*}[h]
    \caption{Statistical properties of the physical and morphological parameters of the full galaxy sample.}
    \centering
    \begin{tabular}{c|c|c|c|c|c|c|c|c|c|c} 
    \hline
     Statistic & \multicolumn{2}{c|}{Mean} & \multicolumn{2}{c|}{Median} & \multicolumn{2}{c|}{$\sigma$} & \multicolumn{2}{c|}{Skewness} & \multicolumn{2}{c}{Kurtosis} \\ \hline
     Parameter & R & D & R & D & R & D & R & D & R & D \\ \hline
    $C$ & 2.66 & 2.65 & 2.63 & 2.62 & 0.22 & 0.22 & 0.92 & 1.10 & 2.18 & 3.33 \\ \hline
    $A$ & 0.02 & 0.01 & 0.01 & 0.01 & 0.06 & 0.06 & 3.30 & 2.22 & 23.36 & 9.32 \\ \hline
    $G$ & 0.49 & 0.49 & 0.49 & 0.49 & 0.04 & 0.03 & 0.47 & 0.46 & 1.25 & 1.00 \\ \hline
    $M_{20}$ & -1.69 & -1.69 & -1.70 & -1.69 & 0.10 & 0.10 & 0.73 & 0.61 & 4.66 & 5.04 \\ \hline
    $n$ & 2.26 & 2.20 & 1.96 & 1.90 & 1.65 & 1.55 & 3.33 & 3.06 & 20.51 & 17.28 \\ \hline
    $\Sigma$ & 8.94 & 8.96 & 9.03 & 9.06 & 0.63 & 0.63 & -0.47 & -0.52 & 0.20 & 0.11 \\ \hline
    $\log(M_*/M_{\odot})$ & 10.30 & 10.30 & 10.35 & 10.35 & 0.50 & 0.50 & -0.36 & -0.36 & -0.27 & -0.28 \\ \hline
    $\log($sSFR/yr$^{-1})$ & -10.94 & -10.88 & -10.80 & -10.80 & 1.00 & 0.94 & -1.12 & -1.19 & 1.65 & 2.37 \\ \hline
    $(g-r)_{\text{norm}}$ & -0.14 & -0.15 & -0.05 & -0.06 & 0.26 & 0.27 & -1.01 & -1.02 & -0.03 & -0.04 \\ \hline
    \end{tabular}
    \label{tab:moments_all_galaxies}
\end{table*}

\section{Discussion}
\label{sec:discussion}

Starting with the mass-size relation, by fitting a single power law to the galaxies ($R_e \propto M_*^{\alpha}$), our results show that the slope for early-type (bulge-dominated) galaxies is steeper than that for late-type (disk-dominated) galaxies in any environment, as was expected \citep[e.g.,][]{Shen+03}. Additionally, this behavior is also observed when we separate galaxies based on their star formation state, consistent with \citet{Chen+24}, in which quiescent galaxies have a steeper slope than star-forming galaxies. When comparing the slopes of the same galaxy types between relaxed and disturbed clusters, we do not find statistically significant differences according to bootstrap analysis. If we consider relaxed clusters to be in a more advanced evolutionary state than disturbed clusters, this could suggest that the evolution of galaxies in terms of mass and size occurs on shorter timescales than cluster relaxation and might therefore be dominated by other factors.

Similarly, the RCS of relaxed and disturbed galaxy clusters does not show statistically significant differences in any of its parameters. Again, this could indicate that galaxy evolution (reflected in the position of galaxies on the CMDs) occurs on timescales that do not strongly depend on the dynamical state of the clusters but rather on shared characteristics of both relaxed and disturbed clusters at low redshift, such as the high density that can facilitate physical processes such as tidal stripping and ram pressure stripping, or the time that the galaxies have to evolve. This is consistent with the results found by \citet{Aldas+23}, which show that for clusters at $z < 0.5$, there are no differences in the colors of galaxy populations. According to the findings of \citet{Pallero+2022}, it is interpreted that at $z > 0.5$, galaxies are quenched in the first cluster they fall into. Therefore, when observing relaxed clusters in that redshift range, the red galaxy population is higher than in disturbed clusters, as galaxies have had more time to evolve in situ. However, at $z < 0.5$, the difference disappears because both relaxed and disturbed clusters have populations that were pre-processed in massive structures beforehand.

Regarding the relationship between morphology and star formation status versus clustercentric distance, we observe that these relations are reproduced consistently, without systematic significant differences between relaxed and disturbed clusters. This is consistent with \citet{Ribeiro+13}, who showed that the fraction of red galaxies as a function of the clustercentric distance does not depend on the evolutionary stage of the systems at low redshifts. However, it is interesting to note that the fraction of disk-dominated galaxies is slightly higher than the fraction of star-forming galaxies, which could suggest that, regardless of the dynamical state of the clusters, galaxy quenching occurs before morphological transformations, or at least on a shorter timescale (if they started at the same time). This behavior is also observed in the morphology-density and SF-density relations, and it is consistent with the results of \citet{Liu+19}.

Concerning the comparison of the physical and morphological parameters as function of the dynamical states, we note that while the KS test reveals statistically significant differences in the distribution of these parameters between relaxed and disturbed clusters, it does not provide insights into the specific nature of these differences, such as whether galaxies are more concentrated or exhibit different trends in one cluster type over the other.

To address this, we analyzed the statistical moments of the distributions. The corresponding values are listed in Table \ref{tab:moments_all_galaxies} for the full galaxy sample, and in the Appendix \ref{sec:appendix-statistical-properties} (Tables \ref{tab:moments_lm_galaxies} to \ref{tab:moments_rs_galaxies}) for each subpopulation separately. Furthermore, the relevant plots are available in the Appendix \ref{sec:appendix-galaxy-properties}. Overall, the central values (mean and median) and dispersion of all physical and structural parameters of the galaxies are very similar when comparing relaxed and disturbed clusters. However, the third (skewness) and fourth (kurtosis) order moments show significant differences in some cases, consistent with the results of the KS tests. Excluding the asymmetry parameter\footnote{We omit the analysis of asymmetry in this discussion because, with our imaging, we push the spatial resolution to the limit of reliability for calculating this parameter.}, we observe a tendency in disturbed clusters toward galaxies being more concentrated (higher $C$), more irregular (higher $M_{20}$), and having higher star formation activity (higher sSFR) compared to relaxed clusters. The increased star formation activity in disturbed clusters is consistent with findings by \citep{Stroe+17}. However, with our results, this trend is weak and only observed in the third- and fourth-order moments.

As is shown in Fig. \ref{fig:parameters-distribution-comparison}, the impact of the cluster's dynamical state on galaxy properties is differential, affecting some populations more strongly than others. For example, high-mass, bulge-dominated, and GV galaxies show minimal or no significant differences in their physical and morphological parameter distributions between relaxed and disturbed clusters. In contrast, low-mass galaxies and galaxies in the red sequence display clearer differences. The parameters showing significant differences in the global comparison (i.e., considering all galaxies together) generally mirror those observed specifically in low-mass and red sequence galaxies, with the exception of the concentration parameter in the latter group. This might indicate that high-mass galaxies are more resilient to environmental disturbances associated with cluster dynamics, while low-mass galaxies are more vulnerable. Furthermore, the absence of differences in concentration among red sequence galaxies could reflect their advanced evolutionary stage, implying that their structural characteristics have largely stabilized, rendering them less sensitive to dynamical processes at low redshift. These findings suggest that, although overall galaxy properties appear broadly stable across varying dynamical states, subtle but measurable differences arise preferentially among galaxy populations more susceptible to environmental processes.

\section{Conclusions}
\label{sec:conclusions}

From a sample of 87 massive galaxy clusters ($M_{500} \geq 1.5 \times 10^{14}$ M$_{\odot}$) within the redshift range $0.10 < z < 0.35$, with classified dynamical states and member galaxies assigned from \citetalias{VelizAstudillo+2024}, we derived physical and morphological parameters using optical imaging from LS DR10 and DES photometry.

Structural parameters were obtained through parametric and nonparametric methods, allowing us to assign morphological types to galaxies. Stellar masses and sSFRs were estimated via SED fitting using seven filters ($grizY$, $W_1$, and $W_2$). With this information, we analyzed galaxy trends with environment, focusing on the comparison of fundamental relations between relaxed and disturbed clusters, and the distribution of galaxy properties.

Under the framework of this study, our results show that the dynamical state of galaxy clusters does not significantly affect their fundamental relations, such as the mass-size relation, the morphology-density relation, the star formation-density relation, or the CMD. However, differences emerge in the distributions of physical and structural parameters, primarily at the level of third- and fourth-order moments, with low-mass and red sequence galaxies being the most impacted.

These findings suggest that, at low redshift, the fundamental relations of massive clusters are already well established and resilient to ongoing dynamical activity. Nevertheless, the presence of subtle but measurable differences in higher-order moments indicates that the dynamical state can still influence the detailed distribution of galaxy properties. This is consistent with a scenario in which the main characteristics of galaxies were shaped earlier during pre-processing in prior structures, with subsequent dynamical activity within clusters leaving residual imprints on the most vulnerable populations.

\begin{acknowledgements}
      We thank the anonymous referee for their valuable and constructive comments, which significantly improved the quality of this manuscript. SVA acknowledges support from the FONDECYT Regular grant 1212046. JLNC \& SVA  acknowledges the financial support of DIDULS/ULS, through the Proyecto Apoyo de Tesis de Postgrado N° PTE2353858. ERC acknowledges the support of the International Gemini Observatory, a program of NSF NOIRLab, which is managed by the Association of Universities for Research in Astronomy (AURA) under a cooperative agreement with the U.S. National Science Foundation, on behalf of the Gemini partnership of Argentina, Brazil, Canada, Chile, the Republic of Korea, and the United States of America.
\end{acknowledgements}

\bibliographystyle{aa}
\bibliography{bibliography}

\onecolumn

\begin{appendix}

\section{Distribution of morphological and physical properties of galaxies}
\label{sec:appendix-galaxy-properties}

In this appendix section, we present the distributions of the physical and morphological parameters of the member galaxies analyzed in this work.
The parameters are shown separately for different galaxy populations: all galaxies, low-mass galaxies, high-mass galaxies, disk-dominated galaxies, bulge-dominated galaxies, BC galaxies, GV galaxies, and red sequence galaxies.
Each plot compares the distributions of galaxies residing in relaxed and disturbed clusters.

\begin{figure*}[h]
    \centering
    \includegraphics[width=1\linewidth]{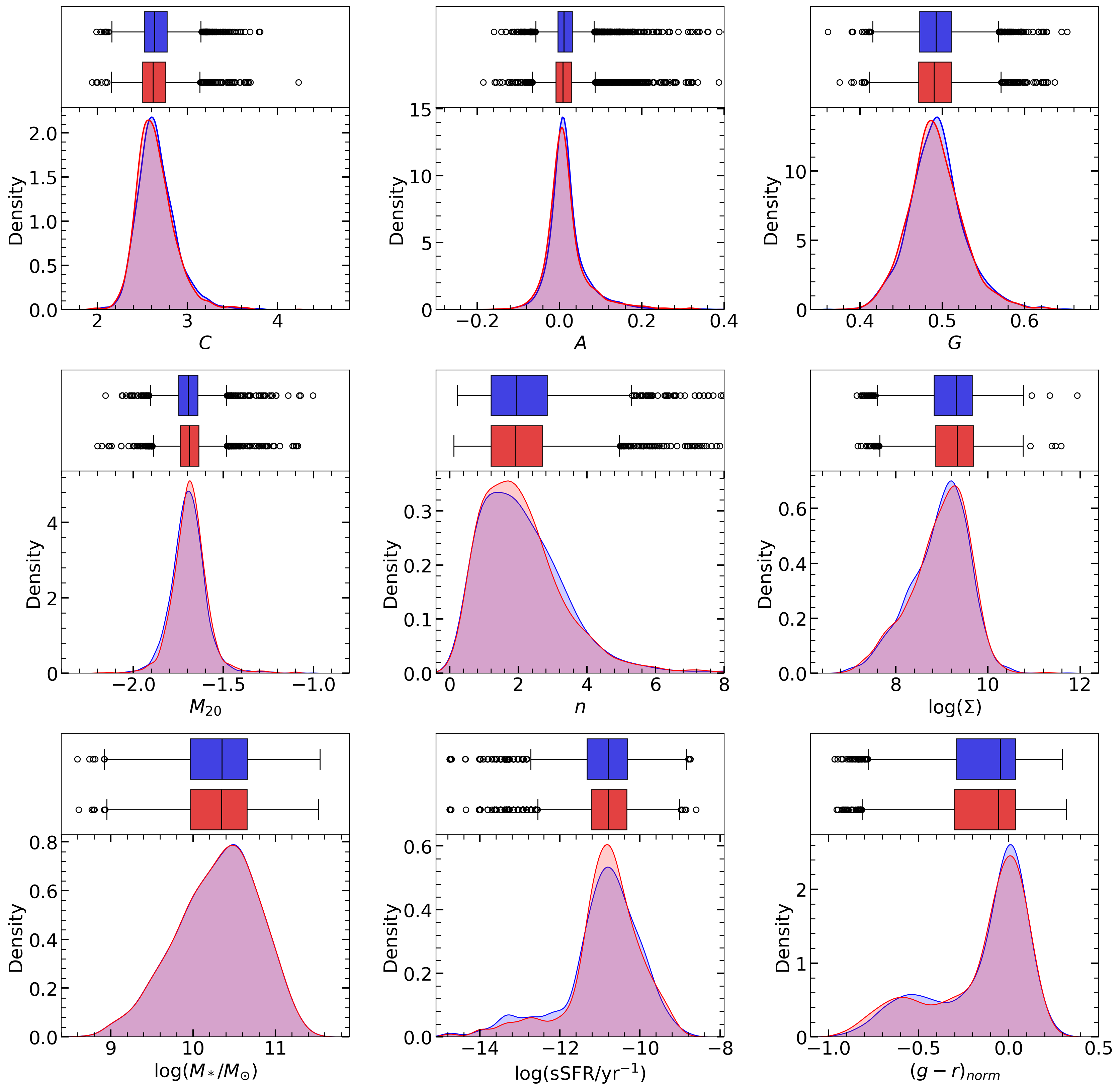}
    \caption{Distribution of the morphological and physical properties for all galaxies in the two environments studied (i.e., relaxed and disturbed galaxy clusters). In all upper panels, boxplots are presented, and in the lower panels, the densities of the distributions estimated with KDE are shown. The colors blue and red represent the relaxed and disturbed galaxy clusters, respectively.}
    \label{fig:distribution_all_galaxies}
\end{figure*}
\FloatBarrier

\begin{figure*}[h]
    \centering
    \includegraphics[width=1\linewidth]{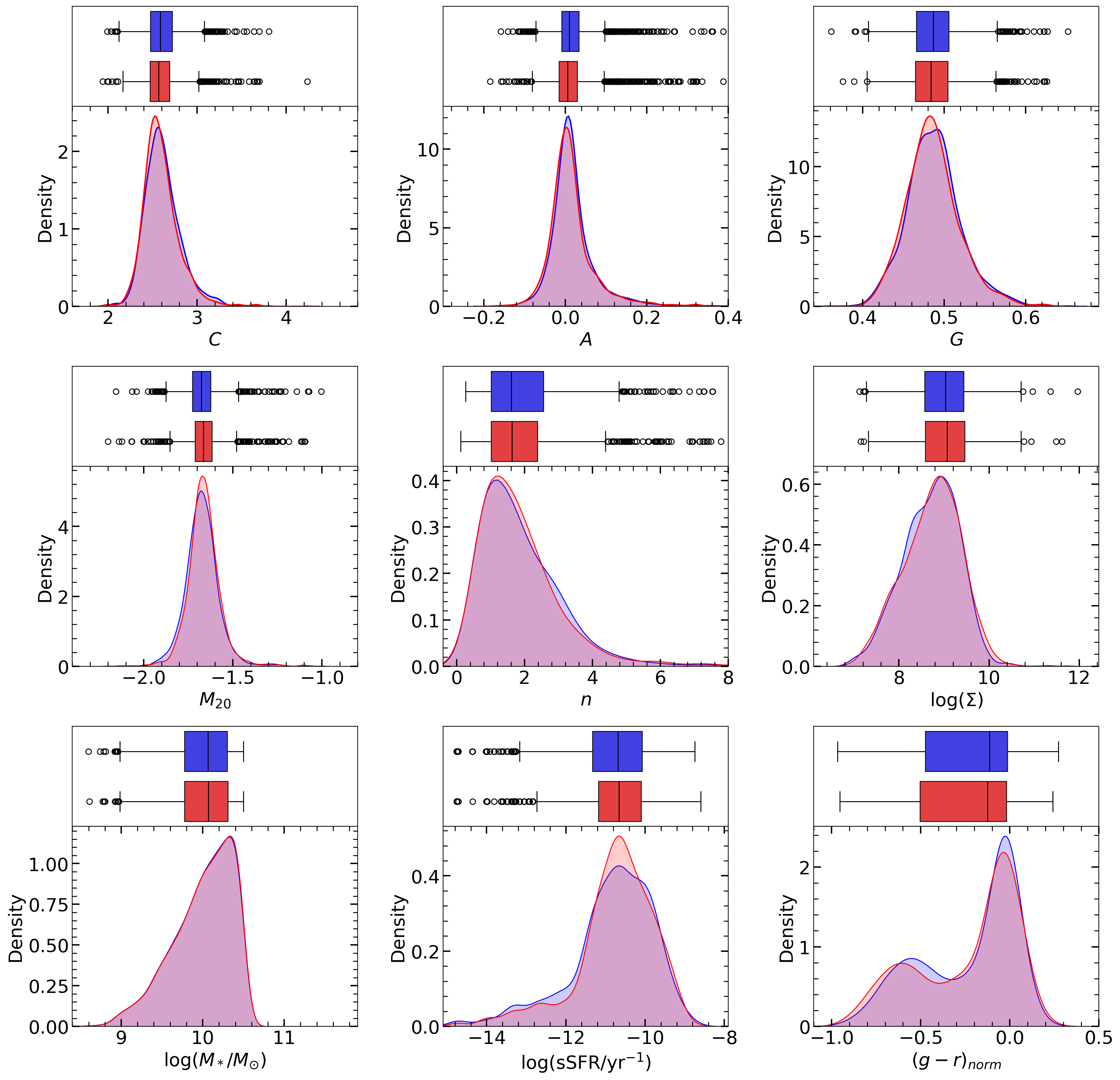}
    \caption{Same as Fig. \ref{fig:distribution_all_galaxies}, but only for low-mass galaxies.}
    \label{fig:distribution_lm_galaxies}
\end{figure*}
\FloatBarrier

\begin{figure*}[h]
    \centering
    \includegraphics[width=1\linewidth]{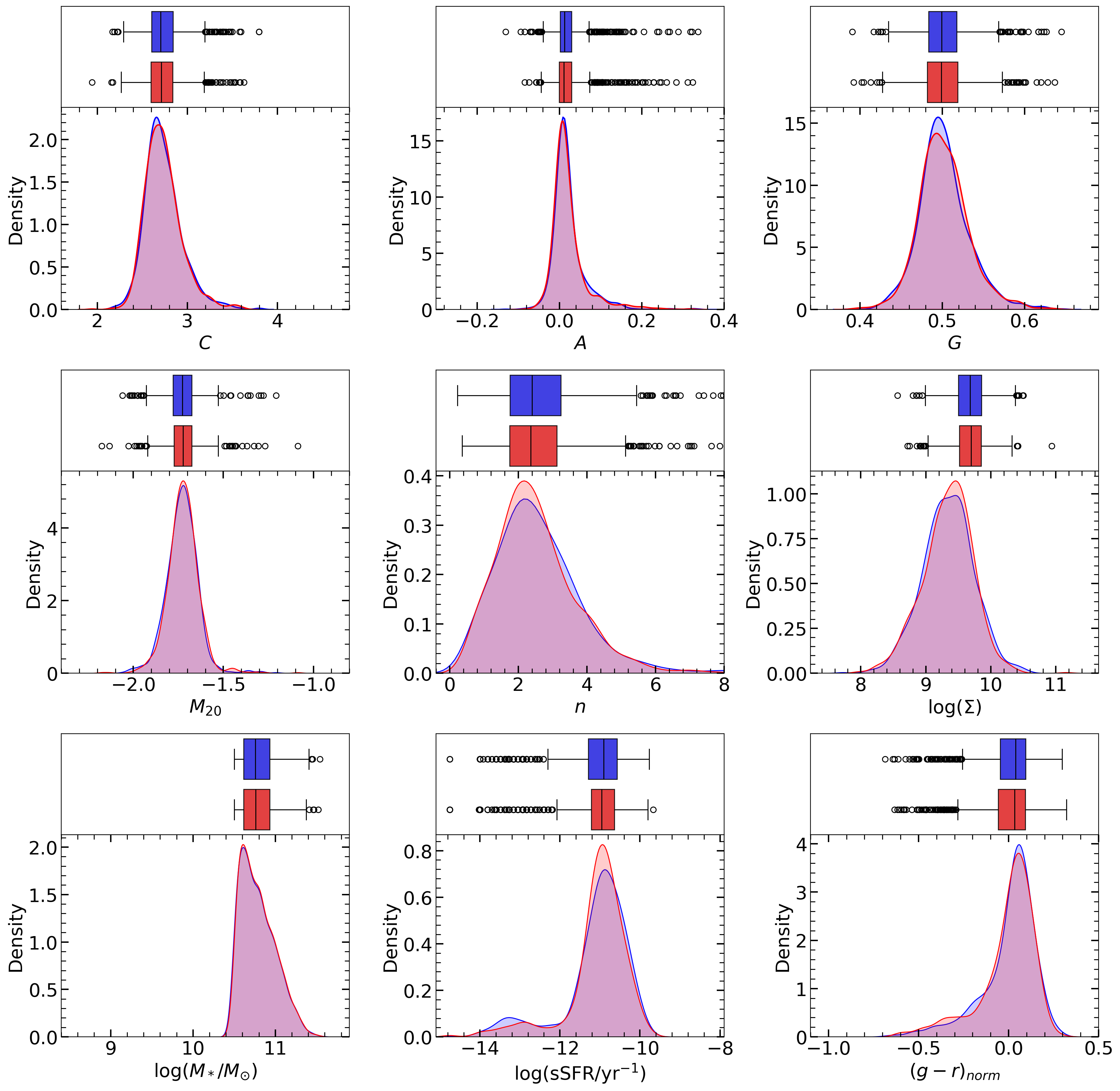}
    \caption{Same as Fig. \ref{fig:distribution_all_galaxies}, but only for high-mass galaxies.}
    \label{fig:distribution_hm_galaxies}
\end{figure*}
\FloatBarrier

\begin{figure*}[h]
    \centering
    \includegraphics[width=1\linewidth]{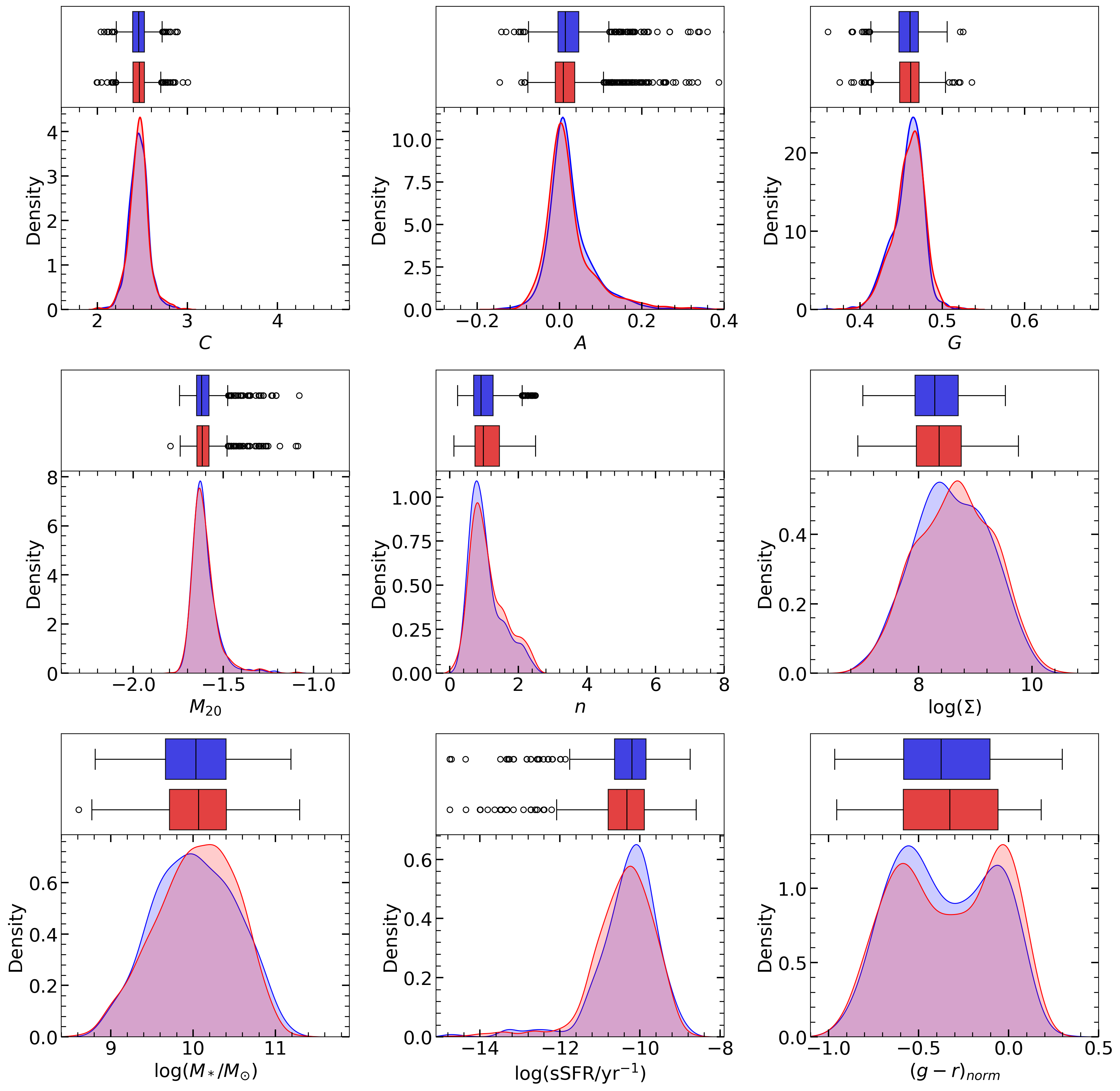}
    \caption{Same as Fig. \ref{fig:distribution_all_galaxies}, but only for disk-dominated galaxies.}
    \label{fig:distribution_disk_galaxies}
\end{figure*}
\FloatBarrier

\begin{figure*}[h]
    \centering
    \includegraphics[width=1\linewidth]{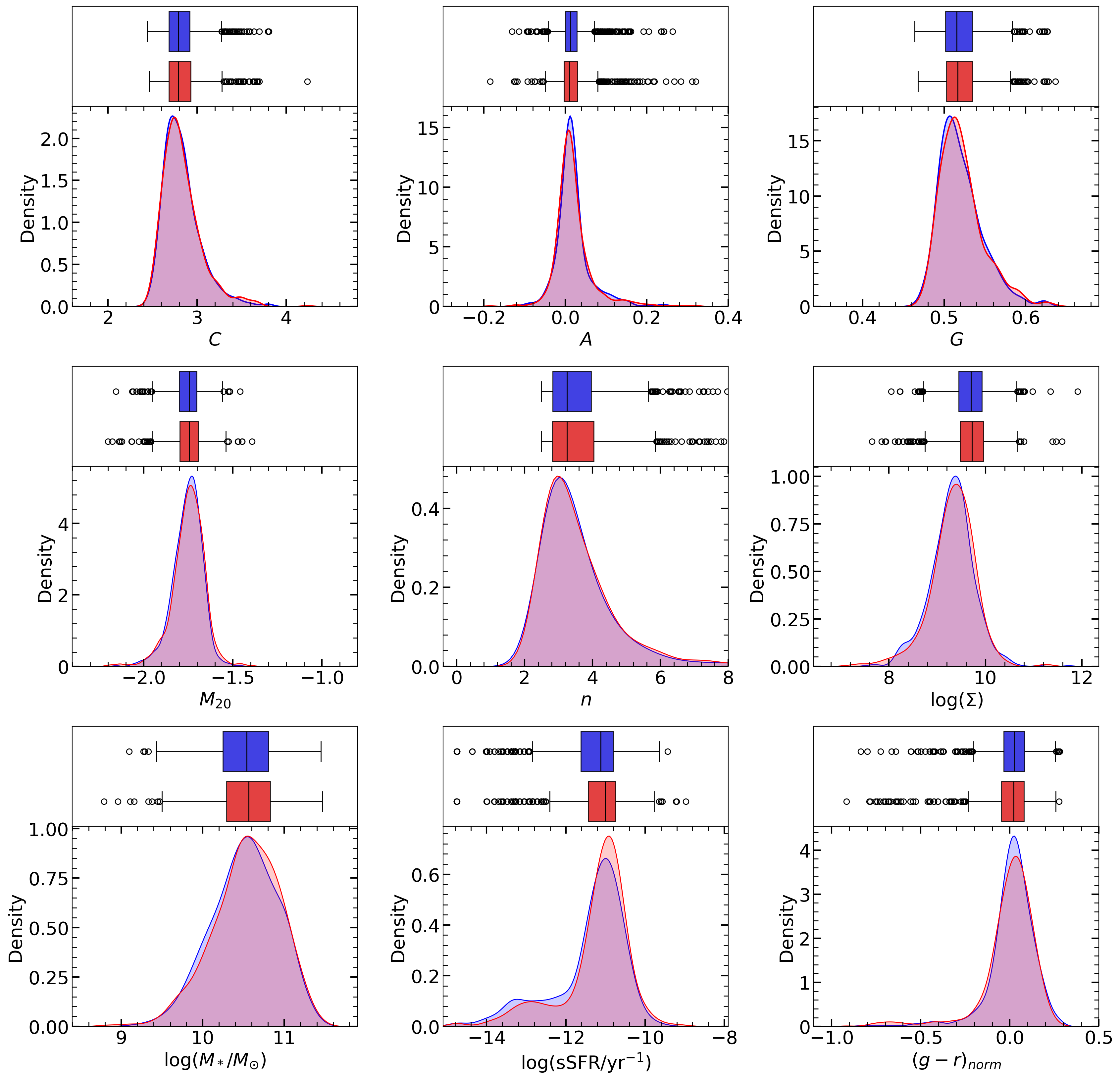}
    \caption{Same as Fig. \ref{fig:distribution_all_galaxies}, but only for bulge-dominated galaxies.}
    \label{fig:distribution_bulge_galaxies}
\end{figure*}
\FloatBarrier

\begin{figure*}[h]
    \centering
    \includegraphics[width=1\linewidth]{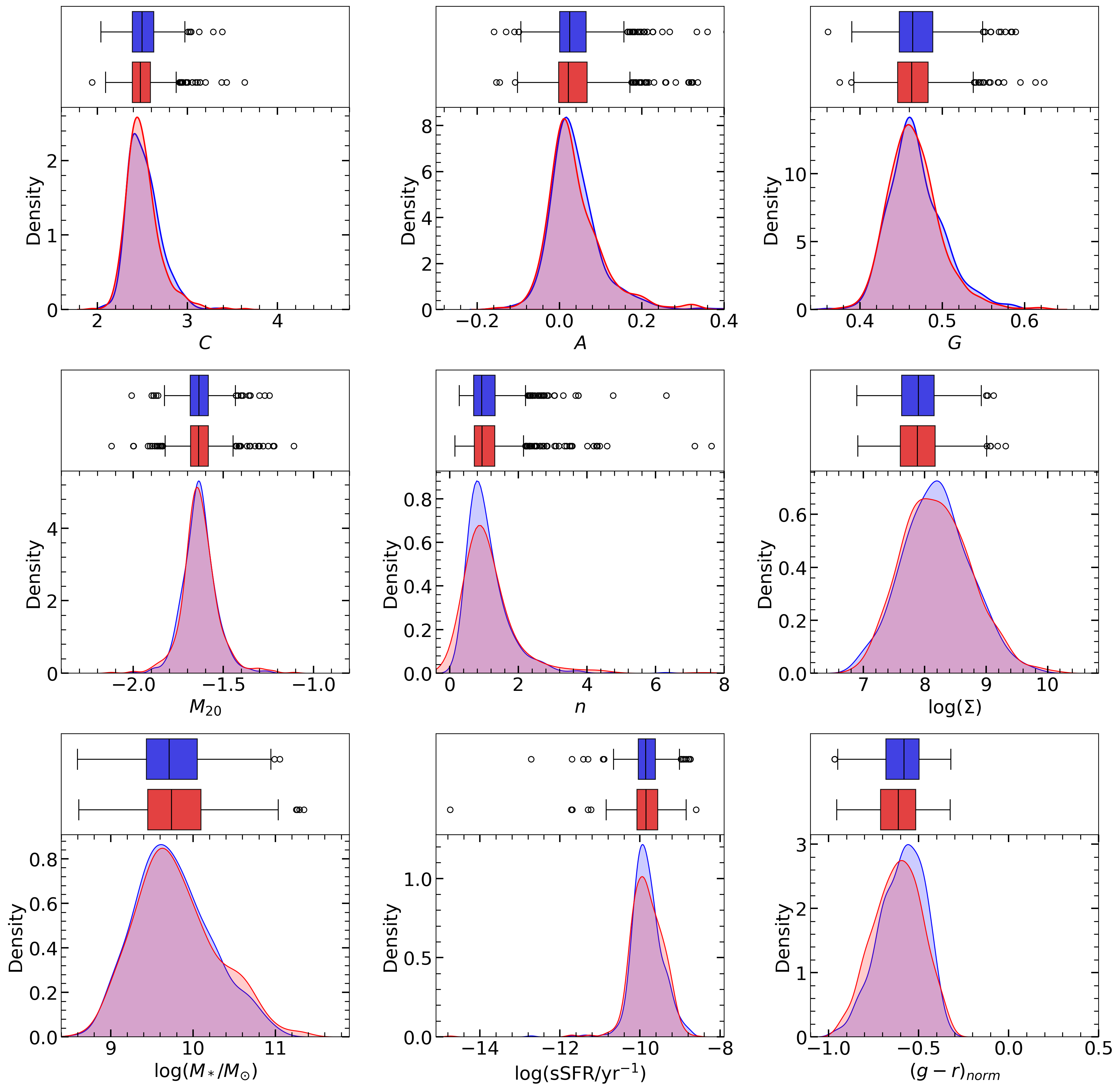}
    \caption{Same as Fig. \ref{fig:distribution_all_galaxies}, but only for BC galaxies.}
    \label{fig:distribution_rs_galaxies}
\end{figure*}
\FloatBarrier

\begin{figure*}[h]
    \centering
    \includegraphics[width=1\linewidth]{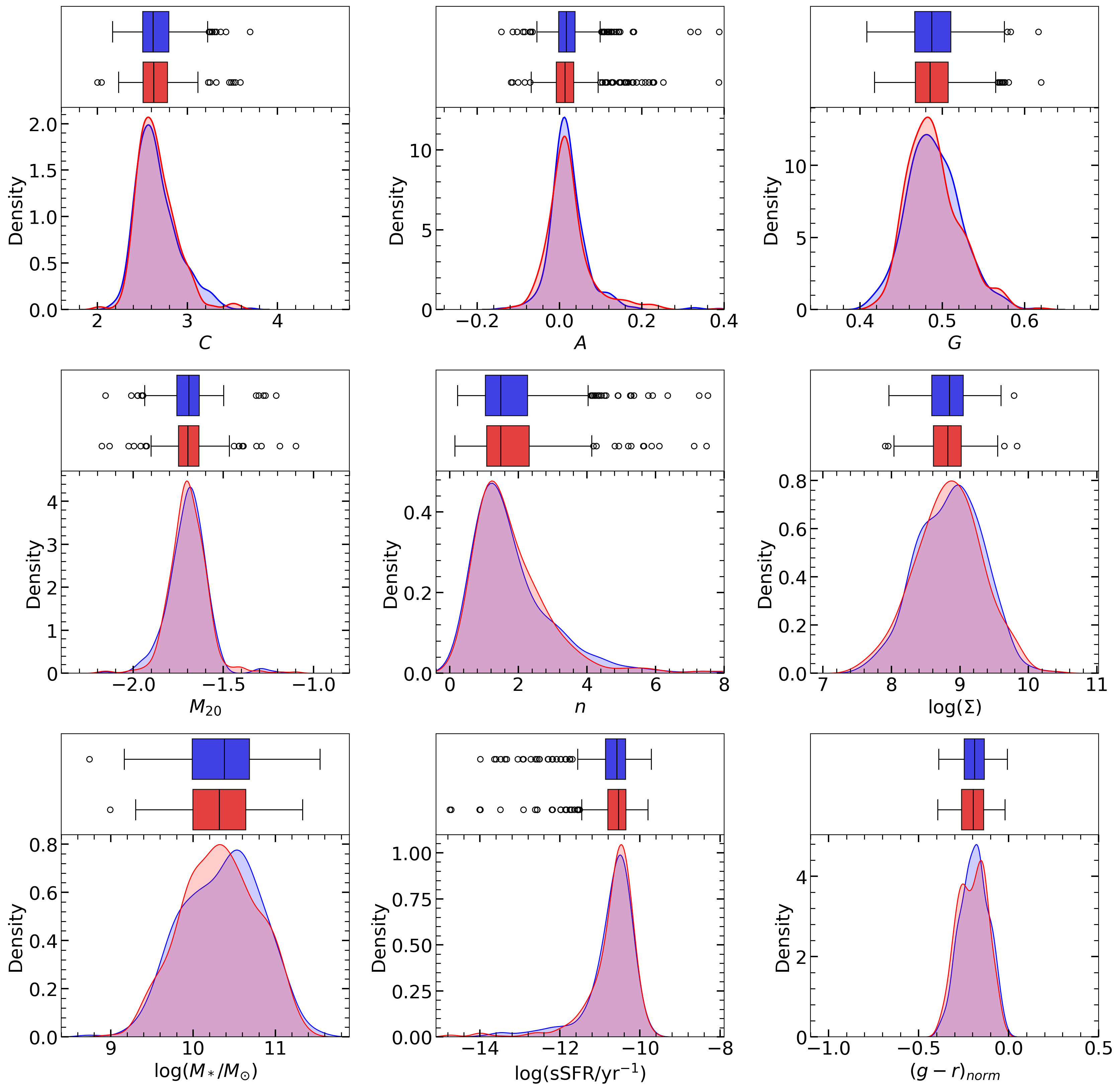}
    \caption{Same as Fig. \ref{fig:distribution_all_galaxies}, but only for GV galaxies.}
    \label{fig:dsitribution_gv_galaxies}
\end{figure*}
\FloatBarrier

\begin{figure*}[h]
    \centering
    \includegraphics[width=1\linewidth]{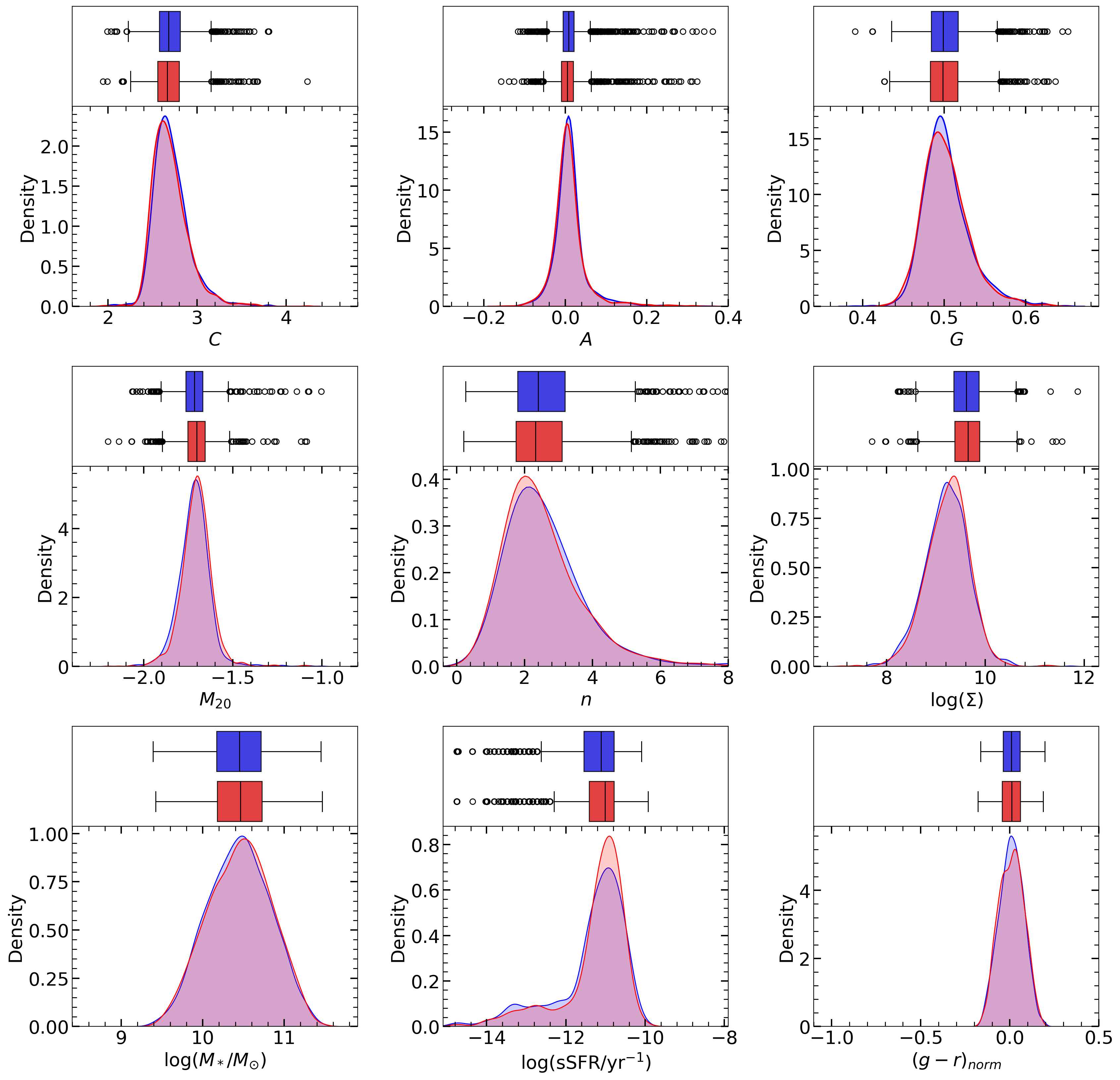}
    \caption{Same as Fig. \ref{fig:distribution_all_galaxies}, but only for red sequence galaxies.}
    \label{fig:dsitribution_rs_galaxies}
\end{figure*}
\FloatBarrier

\section{Statistical properties of the physical and morphological parameters}
\label{sec:appendix-statistical-properties}

This appendix section presents the statistical properties of the physical and morphological parameters of galaxies for all subpopulations defined in this paper. The results are summarized in Tables \ref{tab:moments_lm_galaxies} to \ref{tab:moments_rs_galaxies}.

\begin{table*}[h]
    \caption{Same as Table \ref{tab:moments_all_galaxies}, but for low-mass galaxies.}
    \centering
    \begin{tabular}{c|c|c|c|c|c|c|c|c|c|c} 
    \hline
     Statistic & \multicolumn{2}{c|}{Mean} & \multicolumn{2}{c|}{Median} & \multicolumn{2}{c|}{$\sigma$} & \multicolumn{2}{c|}{Skewness} & \multicolumn{2}{c}{Kurtosis} \\ \hline
     Parameter & R & D & R & D & R & D & R & D & R & D \\ \hline
    $C$ & 2.61 & 2.60 & 2.59 & 2.57 & 0.21 & 0.21 & 1.00 & 1.45 & 2.60 & 2.69 \\ \hline
    $A$ & 0.02 & 0.01 & 0.01 & 0.01 & 0.06 & 0.06 & 3.11 & 2.12 & 20.84 & 8.54 \\ \hline
    $G$ & 0.49 & 0.49 & 0.49 & 0.48 & 0.03 & 0.03 & 0.51 & 0.57 & 1.19 & 1.07 \\ \hline
    $M_{20}$ & -1.67 & -1.66 & -1.68 & -1.67 & 0.10 & 0.10 & 0.86 & 0.66 & 5.59 & 6.00 \\ \hline
    $n$ & 2.02 & 1.99 & 1.61 & 1.63 & 1.66 & 1.66 & 3.75 & 3.67 & 22.82 & 20.80 \\ \hline
    $\Sigma$ & 8.71 & 8.73 & 8.76 & 8.80 & 0.62 & 0.64 & -0.17 & -0.17 & 0.19 & -0.06 \\ \hline
    $\log(M_*/M_{\odot})$ & 10.00 & 10.00 & 10.07 & 10.07 & 0.37 & 0.37 & -0.79 & -0.79 & -0.06 & -0.05 \\ \hline
    $\log($sSFR/yr$^{-1})$ & -10.84 & -10.76 & -10.68 & -10.66 & 1.07 & 1.00 & -1.07 & -1.18 & 1.36 & 1.95 \\ \hline
    $(g-r)_{\text{norm}}$ & -0.23 & -0.24 & -0.11 & -0.12 & 0.27 & 0.28 & -0.65 & -0.66 & -0.87 & -0.88 \\ \hline
    \end{tabular}
    \label{tab:moments_lm_galaxies}
\end{table*}
\FloatBarrier

\begin{table*}[h]
    \caption{Same as Table \ref{tab:moments_all_galaxies}, but for high-mass galaxies.}
    \centering
    \begin{tabular}{c|c|c|c|c|c|c|c|c|c|c} 
    \hline
     Statistic & \multicolumn{2}{c|}{Mean} & \multicolumn{2}{c|}{Median} & \multicolumn{2}{c|}{$\sigma$} & \multicolumn{2}{c|}{Skewness} & \multicolumn{2}{c}{Kurtosis} \\ \hline
     Parameter & R & D & R & D & R & D & R & D & R & D \\ \hline
    $C$ & 2.74 & 2.73 & 2.70 & 2.71 & 0.21 & 0.21 & 1.01 & 0.93 & 2.36 & 2.08 \\ \hline
    $A$ & 0.02 & 0.02 & 0.01 & 0.01 & 0.05 & 0.05 & 3.80 & 2.64 & 28.56 & 9.85 \\ \hline
    $G$ & 0.50 & 0.50 & 0.50 & 0.50 & 0.03 & 0.03 & 0.66 & 0.51 & 1.74 & 1.51 \\ \hline
    $M_{20}$ & -1.73 & -1.72 & -1.73 & -1.72 & 0.09 & 0.09 & 0.54 & 0.57 & 4.12 & 5.70 \\ \hline
    $n$ & 2.66 & 2.54 & 2.40 & 2.36 & 1.56 & 1.26 & 3.14 & 1.72 & 21.60 & 6.75 \\ \hline
    $\Sigma$ & 9.34 & 9.34 & 9.35 & 9.36 & 0.40 & 0.39 & -0.10 & -0.24 & 0.36 & 0.65 \\ \hline
    $\log(M_*/M_{\odot})$ & 10.79 & 10.79 & 10.76 & 10.76 & 0.21 & 0.21 & 0.69 & 0.69 & -0.18 & -0.20 \\ \hline
    $\log($sSFR/yr$^{-1})$ & -11.10 & -11.07 & -10.91 & -10.96 & 0.85 & 0.77 & -1.64 & -1.87 & 2.54 & 4.31 \\ \hline
    $(g-r)_{\text{norm}}$ & 0.00 & -0.01 & 0.04 & 0.03 & 0.15 & 0.16 & -1.46 & -1.49 & 2.57 & 2.27 \\ \hline
    \end{tabular}
    \label{tab:moments_hm_galaxies}
\end{table*}
\FloatBarrier

\begin{table*}[h]
    \caption{Same as Table \ref{tab:moments_all_galaxies}, but for disk-dominated galaxies.}
    \centering
    \begin{tabular}{c|c|c|c|c|c|c|c|c|c|c} 
    \hline
     Statistic & \multicolumn{2}{c|}{Mean} & \multicolumn{2}{c|}{Median} & \multicolumn{2}{c|}{$\sigma$} & \multicolumn{2}{c|}{Skewness} & \multicolumn{2}{c}{Kurtosis} \\ \hline
     Parameter & R & D & R & D & R & D & R & D & R & D \\ \hline
    $C$ & 2.46 & 2.46 & 2.46 & 2.46 & 0.11 & 0.11 & 0.18 & 0.29 & 1.60 & 2.45 \\ \hline
    $A$ & 0.03 & 0.02 & 0.01 & 0.01 & 0.06 & 0.06 & 2.01 & 1.94 & 7.46 & 5.57 \\ \hline
    $G$ & 0.46 & 0.46 & 0.46 & 0.46 & 0.02 & 0.02 & -0.64 & -0.41 & 1.43 & 1.06 \\ \hline
    $M_{20}$ & -1.61 & -1.60 & -1.62 & -1.62 & 0.08 & 0.08 & 2.35 & 2.31 & 9.06 & 8.92 \\ \hline
    $n$ & 1.04 & 1.11 & 0.91 & 0.98 & 0.47 & 0.51 & 1.01 & 0.78 & 0.38 & -0.12 \\ \hline
    $\Sigma$ & 8.58 & 8.63 & 8.56 & 8.66 & 0.65 & 0.67 & -0.03 & -0.07 & -0.61 & -0.61 \\ \hline
    $\log(M_*/M_{\odot})$ & 10.03 & 10.04 & 10.03 & 10.07 & 0.49 & 0.48 & -0.02 & -0.27 & -0.66 & -0.48 \\ \hline
    $\log($sSFR/yr$^{-1})$ & -10.33 & -10.42 & -10.21 & -10.33 & 0.81 & 0.80 & -1.86 & -1.53 & 6.11 & 4.65 \\ \hline
    $(g-r)_{\text{norm}}$ & -0.35 & -0.33 & -0.38 & -0.33 & 0.27 & 0.29 & 0.03 & -0.08 & -1.14 & -1.27 \\ \hline
    \end{tabular}
    \label{tab:moments_disk_galaxies}
\end{table*}
\FloatBarrier

\begin{table*}[h]
    \caption{Same as Table \ref{tab:moments_all_galaxies}, but for bulge-dominated galaxies.}
    \centering
    \begin{tabular}{c|c|c|c|c|c|c|c|c|c|c} 
    \hline
     Statistic & \multicolumn{2}{c|}{Mean} & \multicolumn{2}{c|}{Median} & \multicolumn{2}{c|}{$\sigma$} & \multicolumn{2}{c|}{Skewness} & \multicolumn{2}{c}{Kurtosis} \\ \hline
     Parameter & R & D & R & D & R & D & R & D & R & D \\ \hline
    $C$ & 2.83 & 2.83 & 2.79 & 2.79 & 0.21 & 0.22 & 1.33 & 1.53 & 2.67 & 3.85 \\ \hline
    $A$ & 0.02 & 0.02 & 0.01 & 0.01 & 0.05 & 0.05 & 3.21 & 2.04 & 26.98 & 8.96 \\ \hline
    $G$ & 0.52 & 0.52 & 0.52 & 0.52 & 0.03 & 0.03 & 1.05 & 1.11 & 1.44 & 1.53 \\ \hline
    $M_{20}$ & -1.75 & -1.75 & -1.75 & -1.74 & 0.08 & 0.09 & -0.63 & -0.73 & 1.61 & 3.06 \\ \hline
    $n$ & 3.77 & 3.73 & 3.25 & 3.25 & 1.82 & 1.62 & 4.16 & 3.90 & 23.80 & 22.25 \\ \hline
    $\Sigma$ & 9.31 & 9.32 & 9.33 & 9.35 & 0.46 & 0.48 & -0.04 & -0.63 & 1.66 & 2.76 \\ \hline
    $\log(M_*/M_{\odot})$ & 10.52 & 10.54 & 10.54 & 10.57 & 0.41 & 0.41 & -0.32 & -0.56 & -0.17 & 0.49 \\ \hline
    $\log($sSFR/yr$^{-1})$ & -11.41 & -11.25 & -11.12 & -11.01 & 0.94 & 0.87 & -1.29 & -1.43 & 1.26 & 2.35 \\ \hline
    $(g-r)_{\text{norm}}$ & 0.01 & -0.01 & 0.02 & 0.02 & 0.13 & 0.16 & -2.08 & -2.48 & 8.48 & 8.56 \\ \hline
    \end{tabular}
    \label{tab:moments_bulge_galaxies}
\end{table*}
\FloatBarrier

\begin{table*}[h]
    \caption{Same as Table \ref{tab:moments_all_galaxies}, but for BC galaxies.}
    \centering
    \begin{tabular}{c|c|c|c|c|c|c|c|c|c|c} 
    \hline
     Statistic & \multicolumn{2}{c|}{Mean} & \multicolumn{2}{c|}{Median} & \multicolumn{2}{c|}{$\sigma$} & \multicolumn{2}{c|}{Skewness} & \multicolumn{2}{c}{Kurtosis} \\ \hline
     Parameter & R & D & R & D & R & D & R & D & R & D \\ \hline
    $C$ & 2.52 & 2.51 & 2.50 & 2.48 & 0.18 & 0.19 & 0.87 & 1.46 & 1.53 & 4.49 \\ \hline
    $A$ & 0.04 & 0.04 & 0.02 & 0.02 & 0.07 & 0.07 & 2.57 & 1.59 & 16.22 & 4.45 \\ \hline
    $G$ & 0.47 & 0.47 & 0.46 & 0.46 & 0.03 & 0.03 & 0.70 & 0.95 & 1.01 & 2.34 \\ \hline
    $M_{20}$ & -1.63 & -1.63 & -1.64 & -1.64 & 0.09 & 0.10 & 0.35 & 0.44 & 2.07 & 4.02 \\ \hline
    $n$ & 1.13 & 1.28 & 0.93 & 0.94 & 0.77 & 1.35 & 5.30 & 5.07 & 52.10 & 30.86 \\ \hline
    $\Sigma$ & 8.19 & 8.20 & 8.19 & 8.17 & 0.55 & 0.56 & 0.16 & 0.36 & -0.12 & 0.01 \\ \hline
    $\log(M_*/M_{\odot})$ & 9.76 & 9.80 & 9.71 & 9.74 & 0.46 & 0.49 & 0.38 & 0.49 & -0.26 & -0.10 \\ \hline
    $\log($sSFR/yr$^{-1})$ & -9.83 & -9.83 & -9.86 & -9.86 & 0.39 & 0.44 & -0.88 & -2.88 & 7.16 & 29.21 \\ \hline
    $(g-r)_{\text{norm}}$ & -0.59 & -0.62 & -0.58 & -0.61 & 0.12 & 0.13 & -0.43 & -0.15 & -0.25 & -0.53 \\ \hline
    \end{tabular}
    \label{tab:moments_bc_galaxies}
\end{table*}
\FloatBarrier

\begin{table*}[h]
    \caption{Same as Table \ref{tab:moments_all_galaxies}, but for GV galaxies.}
    \centering
    \begin{tabular}{c|c|c|c|c|c|c|c|c|c|c} 
    \hline
     Statistic & \multicolumn{2}{c|}{Mean} & \multicolumn{2}{c|}{Median} & \multicolumn{2}{c|}{$\sigma$} & \multicolumn{2}{c|}{Skewness} & \multicolumn{2}{c}{Kurtosis} \\ \hline
     Parameter & R & D & R & D & R & D & R & D & R & D \\ \hline
    $C$ & 2.66 & 2.66 & 2.62 & 2.62 & 0.23 & 0.22 & 1.00 & 1.04 & 1.21 & 2.41 \\ \hline
    $A$ & 0.02 & 0.02 & 0.02 & 0.01 & 0.05 & 0.06 & 2.37 & 1.87 & 13.54 & 6.76 \\ \hline
    $G$ & 0.49 & 0.49 & 0.49 & 0.49 & 0.03 & 0.03 & 0.39 & 0.72 & 0.44 & 0.60 \\ \hline
    $M_{20}$ & -1.70 & -1.69 & -1.69 & -1.70 & 0.11 & 0.11 & 0.23 & 0.54 & 3.37 & 5.52 \\ \hline
    $n$ & 1.85 & 1.84 & 1.49 & 1.49 & 1.29 & 1.24 & 2.97 & 2.67 & 16.97 & 11.07 \\ \hline
    $\Sigma$ & 8.85 & 8.83 & 8.89 & 8.85 & 0.47 & 0.59 & -0.08 & -0.06 & -0.24 & -0.03 \\ \hline
    $\log(M_*/M_{\odot})$ & 10.35 & 10.32 & 10.38 & 10.32 & 0.47 & 0.45 & -0.17 & -0.08 & -0.42 & -0.59 \\ \hline
    $\log($sSFR/yr$^{-1})$ & -10.74 & -10.72 & -10.58 & -10.54 & 0.65 & 0.67 & -2.26 & -3.05 & 6.26 & 12.43 \\ \hline
    $(g-r)_{\text{norm}}$ & -0.19 & -0.20 & -0.19 & -0.20 & 0.08 & 0.08 & -0.09 & -0.09 & -0.53 & -0.77 \\ \hline
    \end{tabular}
    \label{tab:moments_gv_galaxies}
\end{table*}
\FloatBarrier

\begin{table*}[h]
    \caption{Same as Table \ref{tab:moments_all_galaxies}, but for red sequence galaxies.}
    \centering
    \begin{tabular}{c|c|c|c|c|c|c|c|c|c|c} 
    \hline
     Statistic & \multicolumn{2}{c|}{Mean} & \multicolumn{2}{c|}{Median} & \multicolumn{2}{c|}{$\sigma$} & \multicolumn{2}{c|}{Skewness} & \multicolumn{2}{c}{Kurtosis} \\ \hline
     Parameter & R & D & R & D & R & D & R & D & R & D \\ \hline
    $C$ & 2.71 & 2.70 & 2.68 & 2.66 & 0.20 & 0.20 & 1.17 & 1.41 & 3.54 & 4.87 \\ \hline
    $A$ & 0.01 & 0.01 & 0.01 & 0.01 & 0.05 & 0.05 & 4.22 & 2.75 & 33.26 & 14.98 \\ \hline
    $G$ & 0.50 & 0.50 & 0.50 & 0.50 & 0.03 & 0.03 & 1.05 & 0.92 & 2.48 & 1.82 \\ \hline
    $M_{20}$ & -1.71 & -1.70 & -1.71 & -1.70 & 0.09 & 0.09 & 1.31 & 0.63 & 8.99 & 7.13 \\ \hline
    $n$ & 2.76 & 2.63 & 2.41 & 2.32 & 1.73 & 1.54 & 3.63 & 3.51 & 21.68 & 22.01 \\ \hline
    $\Sigma$ & 9.22 & 9.24 & 9.23 & 9.26 & 0.44 & 0.44 & 0.08 & -0.18 & 1.08 & 1.54 \\ \hline
    $\log(M_*/M_{\odot})$ & 10.44 & 10.45 & 10.45 & 10.46 & 0.38 & 0.38 & -0.04 & -0.09 & -0.45 & -0.52 \\ \hline
    $\log($sSFR/yr$^{-1})$ & -11.35 & -11.27 & -11.11 & -11.01 & 0.89 & 0.82 & -1.51 & -1.74 & 1.91 & 2.81 \\ \hline
    $(g-r)_{\text{norm}}$ & 0.01 & 0.01 & 0.01 & 0.01 & 0.07 & 0.07 & -0.03 & 0.00 & -0.41 & -0.61 \\ \hline
    \end{tabular}
    \label{tab:moments_rs_galaxies}
\end{table*}
\FloatBarrier

\section{Comparison between LS and DES photometry}
\label{sec:appendix-comparison-between-surveys}

Given that we used photometric data from LS DR10 to derive morphological parameters, fit the red sequences, and apply magnitude cuts, while DES photometry was used for the SED fitting and derivation of physical properties, we compared the photometry from both surveys for the galaxies in our sample.

\begin{figure*}[h]
    \centering
    \includegraphics[width=1\linewidth]{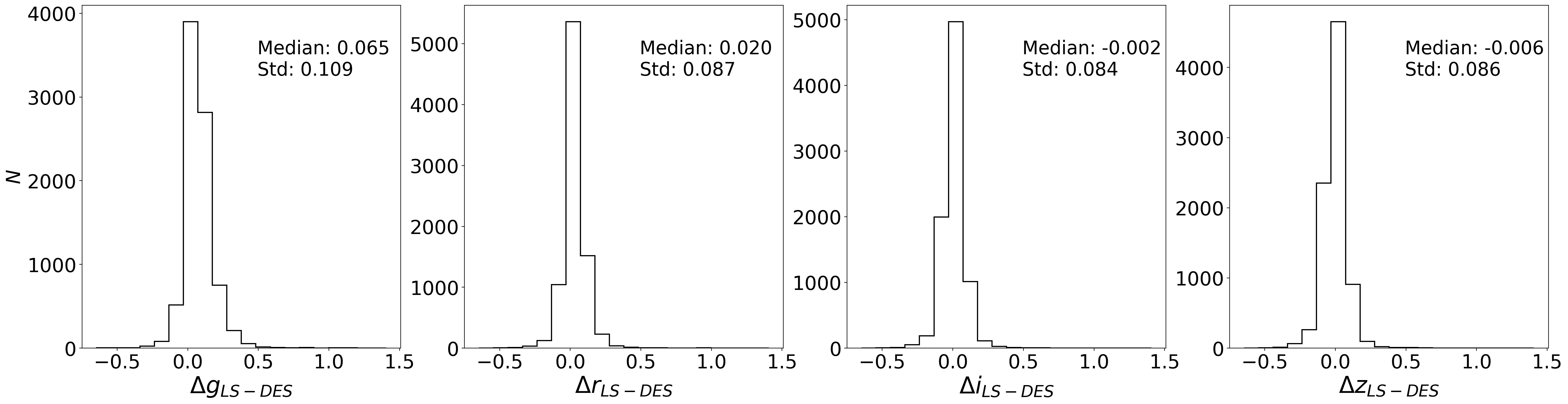}
    \caption{Histograms of the magnitude residuals between LS and DES photometry ($\Delta = m_{\text{LS}} - m_{\text{DES}}$) for the 8,412 galaxies in our sample that passed the quality cuts. Each panel corresponds to one photometric band: $g$ (top left), $r$ (top right), $i$ (bottom left), and $z$ (bottom right). The median and standard deviation of the residuals in each band are indicated within the corresponding panel. The small offsets confirm the overall consistency between both surveys.}
    \label{fig:magnitude-residuals}
\end{figure*}
\FloatBarrier

As shown in Fig. \ref{fig:magnitude-residuals}, the magnitude residuals between the two surveys are small, and we verified that our results are not significantly affected by these differences. Although some numerical values of derived parameters (e.g., the slopes and intercepts of the stacked red sequences and the mass–size relations) vary slightly (Table \ref{tab:comparing-parameters-by-survey}), these shifts are consistent with the small photometric differences.
Importantly, these variations do not impact the main conclusions of this study.

\begin{table*}[h]
    \caption{Values of the slopes of the mass-size relations, and the slopes, intercepts and scatters of the stacked red sequences, obtained using magnitudes from DES and from the LS.}
    \label{tab:comparing-parameters-by-survey}
    \centering
    \begin{tabular}{lccccc} 
    \hline
    Relation & Parameter & \multicolumn{2}{c}{DES value} & \multicolumn{2}{c}{LS value}
    \\ 
    & & Relaxed & Disturbed & Relaxed & Disturbed \\ \hline
    Mass-size (bulge) & Slope & $0.250 \pm 0.018$ & $0.263 \pm 0.020$ & $0.249 \pm 0.018$ & $0.261 \pm 0.020$ \\
    Mass-size (disk) & Slope & $0.027 \pm 0.017$ & $-0.007 \pm 0.017$ & $0.029 \pm 0.018$ & $-0.002 \pm 0.018$ \\
    Mass-size (quiescent) & Slope & $0.224 \pm 0.015$ & $0.209 \pm 0.015$ & $0.239 \pm 0.015$ & $0.209 \pm 0.015$ \\
    Mass-size (star-forming) & Slope & $0.121 \pm 0.018$ & $0.119 \pm 0.017$ & $0.109 \pm 0.018$ & $0.123 \pm 0.018$ \\
    Red sequence & Slope & $-0.044 \pm 0.001$ & $-0.043 \pm 0.001$ & $-0.052 \pm 0.001$ & $-0.055 \pm 0.001$ \\
    Red sequence & Intercept & $0.012 \pm 0.001$ & $0.006 \pm 0.001$ & $0.077 \pm 0.001$ & $0.074 \pm 0.001$ \\
    Red sequence & Scatter & $0.113 \pm 0.001$ & $0.117 \pm 0.001$ & $0.106 \pm 0.001$ & $0.108 \pm 0.001$ \\
    \hline
    \end{tabular}
    \tablefoot{For completeness, errors of 0.001 are quoted for the red sequence parameters; however, as noted in Section \ref{sec:color-magnitude-diagram}, the actual uncertainties are on the order of $10^{-5}$.}
\end{table*}
\FloatBarrier

\end{appendix}

\end{document}